\documentclass[aps,prd,superscriptaddress,twoside,twocolumn,nofootinbib,%
10pt,showpacs,floatfix]{revtex4-1}

\usepackage{amsmath,amssymb}
\usepackage{graphicx,bm}
\usepackage{slashed}
\usepackage{dcolumn}

\allowdisplaybreaks

\begin{document}


\title{Four-quark structure of the excited states of heavy mesons}


\author{Hungchong Kim}%
\email{hungchong@kookmin.ac.kr}
\affiliation{Department of General Education, Kookmin University, Seoul 136-702, Korea}

\author{Myung-Ki Cheoun}%
\email{cheoun@ssu.ac.kr}
\affiliation{Department of Physics, Soongsil University, Seoul 156-743, Korea}

\author{Yongseok Oh}%
\email{yohphy@knu.ac.kr}
\affiliation{Department of Physics, Kyungpook National University, Daegu 702-701, Korea }
\affiliation{Asia Pacific Center for Theoretical Physics, Pohang, Gyeongbuk 790-784, Korea}

\date{\today}


\begin{abstract}

We propose a four-quark structure for some of the excited states of heavy mesons
containing a single charm or bottom quark.
The four-quark wave functions are constructed based on a diquark-antidiquark form
under the constraint that they form an antitriplet $\bar{\bm{3}}_f$ in $\mbox{SU(3)}_f$,
which seems to be realized in some of the excited states listed in Particle Data Group.
Depending on the structure of antidiquark, we construct two possible models for its wave
functions:
Model I) the antidiquark is symmetric in flavor ($\bar{\bm{6}}_f$) and antisymmetric in color
($\bm{3}_c$) and
Model II) the antidiquark is antisymmetric in flavor ($\bm{3}_f$) and symmetric in color
($\bar{\bm{6}}_c$).
To test phenomenological relevance of these wave functions, we calculate the mass differences
among the excited states of spin $J=0,1,2$ using color-spin interactions.
The four-quark wave functions based on Model~I is found to reproduce the observed mass 
of the excited states of heavy mesons.
Also, our four-quark model provides an interesting phenomenology relating to the decay widths
of the excited states.
To further pursue the possibility of the four-quark structure, we make a few predictions for open charm
and open bottom states that may be discovered in future experiments.
Most of them are expected to have broad widths, which would make them difficult to be identified 
experimentally.
However, one resonance with $J=1$ containing bottom and strange quarks is expected to appear
as a sharp peak with its mass around $B^{\bar s}_{1N} \sim 5753$~MeV.
Confirmation of the existence of such states in future experiments will shed light on our
understanding of the structure of heavy meson excited states.

\end{abstract}


\maketitle

\section{Introduction}

Multiquark states, which refer to hadrons composed of four or higher number of quarks, are
very interesting subjects in hadron physics.
Although the ground states of hadrons can be well described by the conventional picture
of quark-antiquark systems for mesons and three-quark systems for baryons, there has been a
controversy over the existence of exotic states including multiquarks and/or glueballs in
hadron spectroscopy.
This is because the conventional quark models taking into account color and flavor degrees of
freedom do not rule out the possible existence of multiquark states.
Indeed, there have been various experiments reporting the candidates of exotic states, which
include $X(3872)$~\cite{Belle03}, $Y(4260)$~\cite{BABAR05b}, and $Z(4430)$~\cite{Belle08b}.
For those mesons, among various interpretations, the four-quark scenarios containing two heavy and
two light quarks are quite promising~\cite{Olsen09,MPPR14,LHCb14}.
Also pentaquark states triggered by the experiments of the LEPS Collaboration at 
SPring-8~\cite{LEPS03} are still under debate both theoretically and experimentally.
Existence of hybrid mesons with gluonic excitations will also be investigated by the Hall-D
experiments at Thomas Jefferson National Accelerator Facility~\cite{Smith02}.

The pure exotic states can be distinguished by their unique quantum numbers but the
existence of crypto-exotic states is hard to identify as their quantum numbers can also be
produced by the conventional pictures of hadrons.
Therefore, some crypto-exotic multiquark states, other than the newly discovered exotic state 
candidates, may already be observed and listed in the current edition of Particle Data Group 
(PDG)~\cite{PDG12}, especially in hadron excited states.
The pioneering work along this direction may be the diquark-antidiquark model advocated by Jaffe
in the 1970s~\cite{Jaffe77a,Jaffe77b}, who proposed the four-quark structure for the scalar
meson nonet, $a_0^{}(980)$,  $f_0^{}(980)$, $\sigma(600)$, and $\kappa(800)$.
(For a review, see Ref.~\cite{Jaffe04}.)
In this model, diquarks, belonging to a color antitriplet and flavor antitriplet having spin 0, are claimed
to be tightly bound and they combine with antidiquarks to form four-quark states.
Thus, the four-quark states constructed in this way are, if there is no orbital excitations, restricted to
have spin zero.
Though this model was confronted with different suggestions based on two-quark picture such as
the $P$-wave ${\bar q} q$~\cite{Torn95} or the mixture of various configurations~\cite{LKO13},
there are other calculations favoring the four-quark picture as well~\cite{MPPR04a,EFG09}.

The lesson from the light quark system certainly provides theoretical motivations for the
possibility of the four-quark structure in the excited states of heavy mesons containing
$c$ or $b$ quark.
Experimentally, the excited states of heavy mesons, which were scarcely explored in the past,
become much richer thanks to recent experimental investigations and
during the last decade or so, the excited states in the open-charm and open-bottom sectors listed
in PDG keep accumulating with various decaying properties.
This can provide a nice environment in investigating the structure of heavy meson excited states.

Indeed, there have been various theoretical investigations for the four-quark structure in the
excited states of open charm mesons.
These include the phenomenological model studies based on the relativistic quark
model~\cite{EFG11}, Glozman-Riska hyperfine interaction~\cite{Jova07},
't Hooft interaction~\cite{Dmit05}, QCD sum rules~\cite{BLMNN05,KO05}, etc.
Even though there are other suggestions based on the two-quark picture~\cite{VR03} or mixing
configurations between two-quark and four-quark states~\cite{VFV06}, it is still worthwhile to pursue
additional signatures for four-quark structure in the excited states of heavy meson systems,
and this is the main motivation of the present investigation.

Our approach for four-quark states is quite phenomenological rather than dynamical.
By closely examining the current data of heavy meson spectroscopy, we will postulate a plausible
flavor structure for the excited states of heavy mesons.
Then possible four-quark wave functions will be constructed accordingly based on a
diquark-antidiquark picture.
Here the diquark is composed of one heavy and one light quark, and the antidiquark is a system
of two light antiquarks.

In the present study, we do not restrict our consideration for the antidiquark state to the scalar type
which belongs to the color triplet and flavor triplet having spin zero.
Instead, we extend our consideration to a more general case by allowing various possible
antidiquark states to see their role in heavy meson excited states.
Based on the observation that the excited heavy meson states listed in PDG have spin 0, 1, 2,
we allow other antidiquark structures other than the scalar state and look for plausible scenarios 
which can accommodate all those spin states within one framework.
To test the phenomenological relevance of various four-quark models generated from this approach, 
the mass differences among heavy mesons will be calculated using color-spin interactions and 
compared with the experimental data.

The paper is organized as follows.
In Sec.~\ref{sec:spectrum}, we examine the excited states of heavy mesons in PDG and
motivate the four-quark picture.
The four-quark wave functions constructed accordingly will be presented in Sec.~\ref{sec:wf}.
After a brief introduction of color-spin interactions in Sec.~\ref{sec:color-spin}, we present our
calculations of the hyperfine masses from the four-quark wave functions in Sec.~\ref{sec:hyperfine}.
Results and discussions are given in Sec.~\ref{sec:results} and we summarize
in Sec.~\ref{sec:summary}.


\begin{table}
\centering
\begin{tabular}{c|ccc|c}  \hline\hline
    & \multicolumn{4}{c}{Lowest-lying states }     \\  \cline{2-5}
Family & Meson  & ~~$I(J^P)$~~ & ~~Mass (MeV)~~ & ~~~$\Gamma$ (MeV) \\ \hline
$D$ & $D^{0}$ & $\frac{1}{2}(0^-)$ & 1864.86 & -   \\
    & $D^{\pm}$ & $\frac12(0^-)$ & 1869.62 & -  \\
    & $D^{*0}$ & $\frac12(1^-)$ & 2006.99 & $ < 2.1$   \\
    & $D^{*\pm}$ & $\frac12(1^-)$ & 2010.29 & 0.096  \\
\hline

$D_s$ &$D_s^{\pm}$ & $0(0^-)$ & 1968.50 & -  \\
      & $D_s^{*\pm}$ & $0(1^-)$ & 2112.3 & $ <  1.9$  \\
\hline

$B$ & $B^{\pm}$ & $\frac12(0^-)$ & 5279.25 & - \\
    & $B^0$ & $\frac12(0^-)$ & 5279.58 & -  \\
    & $B^*$ & $\frac12(1^-)$ & 5325.2 & -  \\
\hline

$B_s$ & $B_s^0$ & $0(0^-)$ & 5366.77 & -  \\
      & $B_s^*$ & $0(1^-)$ & 5415.4 & -  \\  \hline\hline
\end{tabular}
\caption{The lowest-lying resonances with $J^P=0^-,1^-$ in $D$, $D_s$, $B$, $B_s$
families listed in PDG~\cite{PDG12}.}
\label{lowlying}
\end{table}

\section{Heavy meson spectroscopy}
\label{sec:spectrum}

We start with examining $D$ and $B$ meson spectroscopy compiled by the Particle Data Group, 
which motivates the possible four-quark structure for the excited states of heavy mesons.
Listed in Tables~\ref{lowlying} and \ref{excited} are open charm and open bottom mesons
that can be found in the compilation of PDG~\cite{PDG12}.
The lowest-lying states listed in Table~\ref{lowlying} are found to have negative parity.
Their isospins are either $I=1/2$ or $I=0$, and their spins are 0 or 1.
There are 4 (2) mesons in $D$ ($D_s$) family, 3 (2) mesons in $B$ ($B_s$) family.
The excited states, which refer to the resonances with higher masses, are listed in Table~\ref{excited}.
There are 7 (4) mesons in $D$ ($D_s$) family, 3 (2) in $B$ ($B_s$) family.%
\footnote{Some mesons are not included in this list because their quantum numbers are unknown
and their masses are higher than the states that we are considering in this work.}
The excited states listed in Table~\ref{excited} have interesting features to be noted.
Their parity is positive, which is opposite to the lowest-lying case, isospins of all the resonances
are either $I=1/2$ or $I=0$ as in the lowest-lying states, and their spins are $J=0,1,2$.
Within each family, there is a hierarchy in the mass spectrum, i.e., the mass increases with spin $J$,
namely, $m_{J=0}^{} < m_{J=1}^{} < m_{J=2}^{}$.


\begin{table}[t]
\centering
\begin{tabular}{c|ccc|c}  \hline\hline
Family & \multicolumn{4}{c}{Excited states} \\ \cline{2-5}
    & Meson &~~$I(J^P)$~~ & ~~Mass (MeV)~~ & ~~~$\Gamma$ (MeV)~~~ \\ \hline
$D$ &  $D_0^{*0}$ & $\frac12(0^+)$ & 2318.29 & 267 \\
    & $D_0^{*\pm}$ & $\frac12(0^+)$ & 2403 & 283 \\
    & $D_1^0$ & $\frac12(1^+)?$ & 2421.4 & 27.4 \\
    & $D_1^{\pm}$ & $\frac12(1^+)$ & 2423.2 & 25 \\
    & $D_1^{0}$ & $\frac12(1^+)$ & 2427 & 384 \\
    &  $D_2^{*0}$ & $\frac12(2^+)$ & 2462.6 & 49 \\
    &  $D_2^{*\pm}$ & $\frac12(2^+)$ & 2464.3 & 37 \\
\hline

$D_s$&  $D_{s0}^{*\pm}$ & $0(0^+)$ & 2317.8 & $ < 3.8$ \\
     &  $D_{s1}^{\pm}$ & $0(1^+)$ & 2459.6 & $ < 3.5$ \\
     & $D_{s1}^{\pm}$ & $0(1^+)$ & 2535.12 & 0.92 \\
     & $D_{s2}^{*\pm}$ & $0(2^+)$ & 2571.9 & 17 \\
\hline

$B$ & $B_1^{0}$ & $\frac12(1^+)$ & 5723.5 & - \\
    & $B_2^{*0}$ & $\frac12(2^+)$ & 5743 & 23 \\
    & $B_J^*$ & $?(?^?)$       & 5698 & 128 \\

\hline

$B_s$ & $B_{s1}^0$ & $0(1^+)$ & 5828.7 & - \\
      & $B_{s2}^{*0}$ & $0(2^+)$ & 5839.96 & 1.56 \\
\hline\hline
\end{tabular}
\caption{The low-lying excited states with $J^P=0^+,1^+,2^+$ in $D$,$D_s$, $B$, $B_s$ families
collected from PDG.
According to PDG, the quantum numbers ($I$,$J$,$P$) of most excited mesons are yet
to be confirmed.
$D_1^{\pm} (2423)$, whose $J^P$ is unknown, is assigned to have $J^P=1^+$ in our analysis
because of its similar mass with $D_1^0$.}
\label{excited}
\end{table}

As anticipated, the spectrum of the lowest-lying states is consistent with the conventional
$Q\bar{q}$ picture.
They form an antitriplet in $\mbox{SU(3)}_f$ as one can see from Table~\ref{group}, where the
mesons are regrouped according to their spin and parity $J^P$.
In most cases, there are three mesons for each $J^P$, composed by two members in isodoublet
($I=1/2$) and one member in isosinglet ($I=0$).
The mass splitting $\Delta m$ between $I=1/2$ and $I=0$ members is about 90-100 MeV, which,
though somewhat smaller than the quark mass difference $m_s^{} - m_u^{}$, still supports the
formation of $\bar{\bm{3}}_f$.
The only exception is the $B$-mesons in the $J^P=1^-$ channel where one member in isodoublet
($I=1/2$) is missing.
But the mass splitting between $B^*(5325)$ and $B_s^*(5415)$ is again 90 MeV, which is similar in
magnitude to those of other $\bm{\bar{3}}_f$ multiplets.
Even though one more member is anticipated in this channel, we expect that it would be discovered
soon at current experimental facilities and one can safely claim that the $B$-mesons of $J^P=1^-$
also form $\bar{\bm{3}}_f$.
This antitriplet structure is consistent with the two-quark systems having a charm (or a bottom)
and a light antiquark, namely $c{\bar q}~ (q=u,d,s)$ (or $b{\bar q}$) being in relative $S$-wave state.
The negative parity comes out naturally with this quark composition.


\begin{table*}[t]
\centering
\begin{tabular}{c|c|c|c|c|c} \hline\hline
               & Family &~~$J^P$~~ & ~~$I$~~ & ~~Meson~~  & $~~\Delta m$ (MeV) \\ \hline
Lowest-lying & $D$ or $D_s$ & $0^-$&  $\frac12$  & $D^{\pm}(1870),D^0(1865)$ & \\
states          &  & & 0         & $D_s^{\pm}(1968)$    & 101 \\ \cline{3-6}
                &  & $1^-$ & $\frac12$ & $D^{*\pm}(2010),D^{*0}(2007)$ & \\
                & &       & 0         & $D_s^{*\pm}(2112)$ & 104
                \\ \cline{2-6}
               & $B$ or $B_s$& $0^-$ & $\frac12$ & $B^{\pm}(5279),B^0(5280)$ &  \\
               & &       & 0         & $B_s^0(5367)$         & 87 \\ \cline{3-6}
               & & $1^-$ & $\frac12$ & $B^*(5325)$, ? &  \\
               & &      & 0         & $B_s^*(5415)$ & 90   \\  \hline
Excited        & $D$ or $D_s$& $0^+$ & $\frac12$ & $D_0^{*\pm}(2403), \underline{D_0^{*0}(2318)}$  &   \\
states         & &       & 0         & $D_{s0}^{*\pm}(2318)$     & -0.2 \\ \cline{3-6}
               & & $1^+$ & $\frac12$ & $\underline{D_1^{\pm}(2423)},D_1^{0}(2427), \underline{D_1^0(2421)}$ &  \\
               & &       & 0         & $D_{s1}^{\pm}(2460)$ & 37.3 \\
               & &       & 0         & $D_{s1}^{\pm}(2535)$ & 112.8 \\ \cline{3-6}
               & & $2^+$ & $\frac12$ & $\underline{D_2^{*\pm}(2464)},\underline{D_2^{*0}(2463)}$ &  \\
               & &       & 0         & $D_{s2}^{*\pm}(2572)$ & 108.4  \\ \cline{2-6}
               & $B$ or $B_s$ & $0^+$ & $\frac12$ & ?,? &  \\
               &  &     & 0         & ?         & ?  \\ \cline{3-6}
               &  &$1^+$ & $\frac12$ & $\underline{B_1^0(5724)}$, $B_J^*(5698,?)$ &    \\
               &       & & 0         & $B_{s1}^0(5829)$ & 105.9   \\ \cline{3-6}
               & & $2^+$ & $\frac12$ & $\underline{B_2^{*0}(5743)}$, ? &  \\
               & &      & 0         & $B_{s2}^{*0}(5840)$ & 97   \\
               \hline\hline
\end{tabular}
\caption{$D$,$D_s$ and in $B$,$B_s$ families compiled by the quantum numbers $J^P$.
$\Delta m$ is the mass difference between the $I=1/2$ and $I=0$ members, which shows
that most low-lying resonances in each spin channel form $\bar{\bm{3}}_f$ with mass splitting
around 100 MeV.
For the excited states, since the mass difference between the $I=1/2$ states is not small,
the mass splitting $\Delta m$ is calculated using the underlined members in $I=1/2$ as the
reference point.
The $B_J^*$ meson in Table~\ref{excited} is placed with the question mark in the $J^P=1^+$
channel as its quantum numbers are unknown.
}
\label{group}
\end{table*}

We then speculate the structure of excited states listed in Table~\ref{excited}.
Since these states have a positive parity, one can think about two possible ways to construct 
such states.
The first way is based on the two-quark picture.
Here, the states with positive parity can be constructed by orbitally exciting the lowest-lying states
($\ell=1$).
By combining with the spin of the two-quark $j=(0,1)$, one can generate the total spin
$J=0,1,2$ for positive parity states.
Then the mass splitting among the excited states can be generated by spin-orbit forces.
In particular, one can expect that the mass splitting between $J^P=1^+$ and $J^P=0^+$ members
is expected to be about half of the one between $J^P=2^+$ and $J^P=1^+$ members~\cite{Close}.
We see from Table~\ref{excited} that this expectation works well for $D_0^{*\pm}(2403)$,
$D_1^{*\pm}(2423)$, and $D_2^{\pm}(2464)$ but fails for $D_0^{*}(2318)$, $D_1^{0}(2421)$, and
$D_2 (2463)$.

Another way to construct the positive-parity excited states, which we want to pursue in the
present work, is to make the product of the SU(3)$_f$ singlet of ${\bar q} q$ of negative-parity and
the ground states of $c{\bar q}$ (or $b{\bar q}$).
The resulting states contain four quarks and they obviously form a $\bar{\bm{3}}_f$ in SU(3)$_f$.
Of course, the states constructed in this way are close to the two-meson molecular states.
Motivated by this observation, however, what we want to investigate in this work is the general
features of four-quark resonance states in heavy quark sector.
Thus, a similar approach like the diaquonia model~\cite{Jaffe77a,Jaffe77b,APEFL93}
will be adopted for quantitative estimates.

The present investigation is also motivated by the $\bar{\bm{3}}_f$ structure observed explicitly in
the excited states of Table~\ref{group}.
In the $J^P=2^+$ channel of the `$D$ or $D_s$' family, there are three members, namely,
$D_2^{\pm}(2464)$, $D_2^{*0}(2463)$, and $D_{s2}^{*\pm}(2572)$ with the isospins expected
from the $\bar{\bm{3}}_f$ multiplet.
The mass splitting between $I=1/2$ and $I=0$ members is about 108 MeV, which is similar to
the splitting in the lowest-lying mesons.
Thus, the three resonances in $J^P=2^+$ seem to form a $\bar{\bm{3}}_f$.

In the $J^P=1^+$ channel of the `$D$ or $D_s$' family, $D_1^{\pm}(2423)$, $D_1^{0}(2421)$,
and $D_{s1}^{*\pm}(2535)$ seem to form a $\bar{\bm{3}}_f$ with the mass splitting $\Delta m$
of 113 MeV.
However, there is another state, $D_{s1}^{\pm}(2460)$ of $I=0$, which is hard to be classified
as a member of $\bar{\bm{3}}_f$.
Later we will discuss the importance implied by the existence of this state.
We will find that, in the four-quark picture with $\bar{\bm{3}}_f$,
there are two possible ways to make the spin-1 states, and, after taking care of the mixing
between the two, $D_{s1}^{\pm} (2460)$ fits nicely with the member in the spin-1 channel.

In the $J^P=1^+$ channel from the `$B$ or $B_s$' family, there are three resonances.
Here, the $B^*_J(5698)$ may not be a member of an isodoublet with $B^0_1(5724)$ because of
their large mass difference of 26 MeV.
But its existence as well as its quantum number is not well-established yet.
The other two, $B^0_1(5724)$ and $B^0_{s1}(5829)$, have
mass splitting around 106 MeV, similar to the mass splitting expected from the
structure of $\bar{\bm{3}}_f$.
Also in the $J^P=2^+$ channel from the `$B$ or $B_s$' family, there are only two resonances
with the mass splitting 97 MeV, again similar magnitude expected from $\bar{\bm{3}}_f$.
So even though one member in the isodoublet is missing, the two resonances seem to be
members of $\bar{\bm{3}}_f$.

A somewhat puzzling situation can be seen in the $J^P=0^+$ channel.
In the charm sector, even though we have three resonances, the mass of $D_{s0}^{\pm}(2318)$
is almost similar to that of $D_{0}^{*0}(2318)$.
This shows that the $D_{s0}^{\pm}$ can not be a member of $\bar{\bm{3}}_f$ and it may not be
described by our four-quark model with $\bar{\bm{3}}_f$.
Also $D_0^{*\pm}(2403)$, because of its large mass, may not form an isodoublet with
$D_{0}^{*0}(2318)$.
This observation shows that we may need mixing of various configurations for fully describing
the excited heavy meson states, which is, however, beyond the scope of this work.
In the bottom sector, there are no resonances reported from `$B$ or $B_s$' family in the $
J^P=0^+$ channel.
As we will see later, the resonances belonging to $J^P=0^+$, if they are constructed with our
four-quark picture, are found to have strong components in the pseudoscalar-pseudoscalar decay
channels with low-invariant masses.
Because of this, they can have large decay widths, which make them difficult to be discovered
experimentally.
Indeed, we note that $D_{0}^{*0}(2318)$ has a broad width of 267 MeV and was listed in PDG only
recently.%
\footnote{This resonance was not listed in PDG before 2010.}

In this Section, we have examined the excited states of positive parity listed in PDG, which shows
that there are several reasons to believe that most excited states form $\bar{\bm{3}}_f$ in flavor
space.
Though some resonances are still missing in PDG, this examination motivates us to pursue a
possible four-quark structure based on $\bar{\bm{3}}_f$ for the study of the excited states of heavy
mesons containing a charm or a bottom quark.

\section{Four-quark Wave functions}
\label{sec:wf}

In this Section, we construct four-quark wave functions for the excited mesons
in $D$ and $D_s$ families.
As we have discussed in the previous Section, most excited states of heavy mesons
listed in PDG have positive parity with $I=(0,1/2)$ and $J=(0,1,2)$.
In addition, they seem to have the flavor structure of $\bar{\bm{3}}_f$.
Purely from the phenomenological point of view, these properties can be generated by
multiplying an SU(3) singlet ${\bar q}^i q_i$ to the two-quark systems, $Q {\bar q}^i$ ($q_i=u,d,s$),
where $Q$ stands for a heavy quark.
Therefore, $Q=c$ for $D$ and $D_s$ families and $Q=b$ for $B$ and $B_s$ families.
To construct four-quark resonance states instead of molecular states, we follow
the diquark-antidiquark approach~\cite{Jaffe77a,Jaffe77b} and impose the phenomenological
aspect of the $\bar{\bm{3}}_f$ structure mentioned above.
Such four-quark states can be schematically expressed as $Q q_i{\bar q}^j{\bar q}^i$.
To construct the tetra-quark structure, therefore, the possible flavor, color, and spin configurations
of each diquark should be determined.

As far as flavor is concerned, one can separate the antidiquark into two terms: namely,
symmetric ($\bar{\bm{6}}_f$) and antisymmetric ($\bm{3}_f$) combinations as
\begin{eqnarray}
{\bar q}^j{\bar q}^i &=& {1\over 2}\left({\bar q}^j{\bar q}^i + {\bar q}^i{\bar q}^j\right )
+{1\over 2}\left({\bar q}^j{\bar q}^i - {\bar q}^i{\bar q}^j\right )
\nonumber \\
&\equiv& ({\bar q}^j {\bar q}^i)_+ +({\bar q}^j {\bar q}^i)_- .
\label{antidiquark_separation}
\end{eqnarray}
Since these two combinations are orthogonal to each other, we have two possible flavor wave
functions for four-quark states:
\begin{widetext}
\begin{eqnarray}
\mbox{Case~1:}~ D^{{\bar q}^j} \Big |_{\rm flavor} &=&
{1\over \sqrt{2}}\sum_{q_i=u,d,s} Qq_i ({\bar q}^j {\bar q}^i)_+
= {1\over \sqrt{2}}[Qu({\bar q}^j{\bar u})_+ +Qd({\bar q}^j{\bar d})_+ +Qs({\bar q}^j {\bar s})_+] ,
\label{symmetric flavor wave function} \\
\mbox{Case~2:}~ D^{{\bar q}^j} \Big |_{\rm flavor} &=& \sum_{q_i=u,d,s} Qq_i ({\bar q}^j {\bar q}^i)_-
= Qu({\bar q}^j{\bar u})_- +Qd({\bar q}^j{\bar d})_- +Qs({\bar q}^j {\bar s})_- .
\label{antisymmetric flavor wave function}
\end{eqnarray}
\end{widetext}
Here $Q=c$ so that these wave functions denote the excited states of $D$-mesons.
When ${\bar q}^j={\bar u}$ or ${\bar d}$, these four-quark wave functions may represent
the excited states in $D$ family, and when ${\bar q}^j={\bar s}$, they may be the excited states
in $D_s$ family.
Clearly from this equation, we see that $D^{{\bar q}^j}$ in either case form $\bar{\bm{3}}_f$
separately in flavor space.

In color space, the diqaurk belongs to either $\bar{\bm{3}}_c$ or $\bm{6}_c$ and
the antidiquark to $\bm{3}_c$ or $\bar{\bm{6}}_c$.
Thus, to make colorless four-quark states, the diquark and antidiquark should be in either
$(\bar{\bm{3}}_c, \bm{3}_c)$ or $(\bm{6}_c, \bar{\bm{6}}_c)$.
Possible spins of the diquark and antidiquark, represented by $J_{12}$ and $J_{34}$, respectively,
are $0,1$.
By combining these spins, one can generate the total spin states for the four-quark states
as $J=0,1,2$ since $\bm{J} = \bm{J}_{12} + \bm{J}_{34}$.
Depending on specific flavor combination we choose, we can determine the possible color and
spin configurations.

\subsection{\boldmath Antidiquark: flavor symmetric case $({\bar q}^j {\bar q}^i)_+$}

We first discuss the case when the antidiquark is symmetric in flavor, i.e., $({\bar q}^j {\bar q}^i)_+$.
Since the antidiquark should be totally antisymmetric when spin, flavor, and color are considered all
together, it can be either $\bm{3}_c$ or $\bar{\bm{6}}_c$ in color space.
When it is in $\bm{3}_c$, since this is antisymmetric in color indices, the antidiquark spin is restricted
to $J_{34}=1$ in order to make totally antisymmetric $({\bar q}^j {\bar q}^i)_+$ systems.
On the other hand, the $Qq$ diquark that contains a heavy quark is not constrained by the Pauli
principle.
Thus, if the four-quark state (namely diquark-antidiquark system) has spin zero, possible spin
configuration for the $Qq$ diquark and the $\bar{q} \bar{q}$ antidiquark
is $J_{12}=1$, and $J_{34}=1$, respectively, which we denote as
$|J, J_{12}, J_{34} \rangle = | 0 1 1 \rangle$.
For spin-1 case, we have two spin configurations:
i) $|J, J_{12}, J_{34} \rangle = | 1 0 1 \rangle$ and ii) $|J, J_{12}, J_{34} \rangle = | 1 1 1 \rangle$.
If this situation is realized in the meson spectroscopy, the physical states should be mixing states
of these two states in the $J=1$ channel.
For $J=2$, the only possible spin configuration is $ | J, J_{12}, J_{34} \rangle = | 2 1 1 \rangle$.
Thus, if the four-quark states are constructed under the assumption that the antidiquark is in flavor
symmetric and color antisymmetric state (${\bm 3}_c$), there are one state with $J=0$, two states with
$J=1$, and one state with $J=2$.
These numbers of states are seemingly consistent with the experimental spectra observed
for the $D$ and $D_s$ family as one can see from Table~\ref{group}, suggesting that this
model is promising for the excite states of open charm mesons.

Given the flavor part of the four-quark wave function in Eq.~(\ref{symmetric flavor wave function}),
it is straightforward to incorporate the color part.
Since the diquark (antidiquark) belongs to $\bar{\bm{3}}_c$ ($\bm{3}_c$) in color, we obtain the
four-quark wave function as
\begin{widetext}
\begin{eqnarray}
D_{J}^{{\bar q}^j} [({\bar q}^j {\bar q}^i)_+ \in \bm{3}_c]
&=& {1\over \sqrt{24}} \sum_{q_i=u,d,s}
\left\{ \sum_{a,b,d,e,f} \varepsilon_{abd}^{} \ \varepsilon^{aef} \,
[(Q)^b (q_i)^d]_{J_{12}^{}=0,1}
[\bm{(} ({\bar q}^j)_e({\bar q}^i)_f \bm{)}_+]_{J_{34}^{}=1}
\right\} ,
\label{3 wave function}
\end{eqnarray}
\end{widetext}
where $a$, $b$, $d$, $e$, $f$ are color indices.
The numerical factor $1/\sqrt{24}$ in Eq.~(\ref{3 wave function}) includes the color normalization
$1/\sqrt{12}$ as well as the flavor normalization $1/\sqrt{2}$ from
Eq.~(\ref{symmetric flavor wave function}).
We have also indicated that the $Qq$ diquark can have spin 0 or 1, but the $\bar{q}\bar{q}$
antidiquark in the present configuration can have spin 1 only.

When the antidiquark is in a color symmetric state of $\bar{\bm{6}}_c$, its spin is restricted to an antisymmetric state, i.e., $J_{34}=0$.
Then the possible spin configurations are $ | J, J_{12}, J_{34} \rangle = | 0 0 0 \rangle$ for $J=0$,
and $ | J, J_{12}, J_{34} \rangle = | 1 1 0 \rangle$ for $J=1$.
This model with $\bar{\bm{6}}_c$ cannot generate $J=2$ state and thus this scenario alone cannot
explain the observed excited states whose spins range from 0 to 2.
If one wants to describe all the states with spin 0,1,2 within the same framework, one should construct
a model allowing both color configurations, $\bm{3}_c$ and $\bar{\bm{6}}_c$, for the antidiquark,
since the two configurations can mix each other.
This is the only way that the $\bar{\bm{6}}_c$ configuration can enter into the framework.
With this mixing scheme, however, though we can generate all the spin states, number of states
generated from this scenario seems too many.
There should be two states in spin 0, three states in spin 1, and one state in spin 2, which is
not consistent with the observed excited states.
For example, in Table~\ref{group}, if one counts the number of mesons in $D$ family with charge
zero, there are one meson in spin 0, two mesons in spin 1, and one meson in spin 2.
For charged mesons in $D$ family, there are one meson in spin 0, one in spin 1, and one in spin 2.
Therefore, the mixing scheme requires additional two or three mesons of similar masses to be
discovered in $D$ family, which seems to be inconsistent with the present observations.
Thus, the mixing scheme, allowing both color states ${\bf 3}_c$ and ${\bf\bar 6}_c$ for the
antidiquark, may be implausible for the excited states.
In the present work, when the antidiquark is flavor symmetric $({\bar q}^j {\bar q}^i)_+$, we consider
the antidiquark with the color state ${\bm 3}_c$ only.
This model will be referred to as Model I.
Our discussion on colors and possible spin configurations for diquark and antidiquark, when the
antidiquark is in flavor symmetric state, is summarized in Table~\ref{DS spin configuration}.


\begin{table}[t]
\centering
\begin{tabular}{c|c|c|c|c}  \hline\hline
\multicolumn{2}{c|}{$Q q_i$ }            &  \multicolumn{2}{c|}{$({\bar q}^j {\bar q}^i)_+$}  &
$Q q_i ({\bar q}^j {\bar q}^i)_+ $      \\
\hline
Spin ($=J_{12}$) &  Color    & Spin ($=J_{34})$   &  Color & $|J, J_{12}, J_{34} \rangle $
\\ \hline
0     &  $\bar{\bm{3}}_c$ & 1 &  $\bm{3}_c$      & $ | 1 0 1 \rangle $ \\
1     &  $\bar{\bm{3}}_c$ & 1 &  $\bm{3}_c$      & $ | 0 1 1 \rangle $,
$ | 1 1 1 \rangle $, $ | 2 1 1 \rangle $ \\
\hline
 0       & $\bm{6}_c$ & 0  & $\bar{\bm{6}}_c$      & $ | 0 0 0 \rangle $ \\
1     &  $\bm{6}_c$ & 0 &  $\bar{\bm{6}}_c$      & $ | 1 1 0 \rangle $ \\
\hline\hline
\end{tabular}
\caption{Possible spins and colors of the $Qq$ diquark, the $\bar{q}\bar{q}$ antidiquark,
and four-quark states when the $\bar{q}\bar{q}$ antidiquark is symmetric in flavor,
$({\bar q}^j {\bar q}^i)_+$.
The case with the antidiquark in the color state of ${\bf 3}_c$ is referred to as Model I.}
\label{DS spin configuration}
\end{table}

\subsection{\boldmath Antidiquark: flavor antisymmetric case $({\bar q}^j {\bar q}^i)_-$}

The other flavor configuration of the $\bar{q}\bar{q}$ antidiquark is antisymmetric combination,
$({\bar q}^j {\bar q}^i)_-$.
Again the Pauli principle requires that the antidiquark is antisymmetric when spin, flavor, and color
degrees of freedom are considered all together.
We begin with the color symmetric state $\bar{\bm{6}}_c$.
Since the antidiquark is flavor antisymmetric, its spin state is restricted to the symmetric state
of $J_{34}=1$ in order to make a totally antisymmetric $({\bar q}^j {\bar q}^i)_-$ system.
Since the spin of the $Qq$ diquark can be $J_{12}=0,1$, the spin of the four-quark states
can be $J=0,1,2$.
If the four-quark state (namely diquark-antidiquark system) has spin zero, the possible spins
for the diquark and the antidiquark are $J_{12}=1$, $J_{34}=1$ so that the spin configuration of the
four-quark system is $|J, J_{12}, J_{34} \rangle = | 0 1 1 \rangle$.
When $J=1$, however, we again have two spin configurations:
i) $|J, J_{12}, J_{34} \rangle=|1 0 1 \rangle$ and ii) $ | J, J_{12}, J_{34} \rangle = | 1 1 1 \rangle$.
When $J=2$, the only possible spin configuration is $ | J, J_{12}, J_{34} \rangle = | 2 1 1 \rangle$.
Thus, in this scenario, one can construct one state in spin-0, two states in spin-1, and one state
in spin-2, again seemingly agreeing with the excited meson spectra in charm sector.

The four-quark wave function can be constructed straightforwardly.
Incorporating the color part into Eq.~(\ref{antisymmetric flavor wave function}), we obtain the
four-quark wave functions as
\begin{widetext}
\begin{eqnarray}
D_{J}^{{\bar q}^j} ~[({\bar q}^j {\bar q}^i)_- \in \bar{\bm{6}}_c]
= {1\over \sqrt{24}} \sum_{q_i=u,d,s} ~\sum_{a,b} \Bigl\{
[(Q)_a (q_i)_b]_{J_{12}=0,1}
[(({\bar q}^j)^a({\bar q}^i)^b)_-]_{J_{34}=1}
+ [(Q)_a (q_i)_b]_{J_{12}=0,1}
[(({\bar q}^j)^b({\bar q}^i)^a)_-]_{J_{34}=1} \Bigr \} ,
\nonumber \\
\label{6 wave function}
\end{eqnarray}
\end{widetext}
where the possible spins for the diquark $J_{12}$ and antidiquark $J_{34}$ are indicated explicitly.
Here, the factor $1/\sqrt{24}$ comes from the color part.

When the antidiquark is in a color antisymmetric state of $\bm{3}_c$, since we are considering
flavor antisymmetric wave function for the antidiquark, its spin is restricted to an
antisymmetric state, i.e., $J_{34}=0$.
With this constraint, the possible spin configurations are $|J, J_{12}, J_{34} \rangle = | 0 0 0 \rangle$
for $J=0$, and $ | J, J_{12}, J_{34} \rangle = | 1 1 0 \rangle$ for $J=1$.
We can not generate the spin 2 state in this configuration.
Here we have similar situation discussed in the last part of the previous subsection.
With a similar argument, this scheme with ${\bm 3}_c$, even if we allow the mixing among the
$\bm{3}_c$ and $\bar{\bm{6}}_c$ cases, may not be relevant for the excited states.
In this work, when the antidiquark is flavor antisymmetric $({\bar q}^j {\bar q}^i)_-$, we consider
the color state with $\bar{\bm{6}}_c$ only.
This model is referred to as Model II from now on.
Our discussion on colors and possible spin configurations for $Qq$ diquark and $\bar{q} \bar{q}$
antidiquark, when the antidiquark is in flavor antisymmetric state, is summarized in
Table~\ref{DA spin configuration}.


\begin{table}[t]
\centering
\begin{tabular}{c|c|c|c|c}  \hline\hline
\multicolumn{2}{c|}{$Q q_i$ }            &  \multicolumn{2}{c|}{$({\bar q}^j {\bar q}^i)_-$}  &
$Q q_i ({\bar q}^j {\bar q}^i)_- $      \\
\hline
Spin ($=J_{12}$) &  Color    & Spin ($=J_{34})$   &  Color & $ | J, J_{12}, J_{34} \rangle $  \\ \hline
0     &  $\bar{\bm{3}}_c$ & 0 &  $\bm{3}_c$      & $|0 0 0 \rangle $ \\
1     &  $\bar{\bm{3}}_c$ & 0 &  $\bm{3}_c$      & $|1 1 0 \rangle $ \\
\hline
 0       & $\bm{6}_c$ & 1  & $\bar{\bm{6}}_c$      & $|1 0 1 \rangle $ \\
1     &  $\bm{6}_c$ & 1 &  $\bar{\bm{6}}_c$      & $|0 1 1 \rangle $, $
|1 1 1 \rangle $, $|2 1 1 \rangle $ \\
\hline\hline
\end{tabular}
\caption{Possible spins (and colors) of the $Qq$ diquark, the $\bar{q}\bar{q}$ antidiquark,
and four-quark states when the antidiquark is antisymmetric in flavor, $({\bar q}^j {\bar q}^i)_-$.
The case with the antidiquark in the color state of $\bar{\bm{6}}_c$  is referred to as Model II. }
\label{DA spin configuration}
\end{table}

\section{Color-spin interactions}
\label{sec:color-spin}

To test the four-quark wave functions constructed in the previous Section, we now use the color-spin
interaction to estimate the mass splittings among heavy mesons of our concern.
The color-spin interaction takes the following
simple form~\cite{LY09,Keren07,OT89,Silve92,GR81}
\begin{equation}
V = \sum_{i < j} v_0^{}\, \lambda_i \cdot \lambda_j \, \frac{J_i\cdot J_j}{m_i^{} m_j^{}} ,
\label{color_spin}
\end{equation}
when the spatial dependence is integrated out.
Here $\lambda_i$ denotes the Gell-Mann matrix, $J_i$ the spin,
and $m_i^{}$ the constituent mass of the $i$-th quark.
The overall strength of the color-spin interaction is controlled by the parameter $v_0^{}$, which
needs to be determined from the experimental data.
This interaction is basically a generalization of the dipole-dipole electromagnetic interaction to take
into account effectively the gluon exchange among constituent quarks.

Using the color-spin interaction, the hadron mass can be calculated by
\begin{equation}
M_H \sim \sum_{i} m_i^{} + \langle V \rangle  ,
\label{mass}
\end{equation}
where the hyperfine mass $\langle V \rangle$ is obtained by using an appropriate hadron
wave function.
A nice aspect of this approach is that, even though Eq.~(\ref{mass}) is not precise enough to
reproduce the experimental masses, the mass differences among hadrons are successfully
explained by the differences in the hyperfine masses,
\begin{equation}
\Delta M_H \sim \Delta \langle V \rangle .
\label{mass gap}
\end{equation}


\begin{table}[t]
\centering
\begin{tabular}{c|c|c}  \hline\hline
              & $\Delta m$ from data & $\Delta m$ from $\langle V \rangle$ \\ \hline
$\Delta -N$        & 292                 & 292 (fit) \\
$\Sigma -\Lambda$        & 77                  & 66.2 \\
$\Sigma^* -\Sigma   $    & 192                 & 192.7 \\
$\Xi^* -\Xi$             & 211                 & 192.7 \\ \hline
$\Sigma_c -\Lambda_c$    & 167                 & 151.8 \\
$\Sigma_c^* -\Sigma_c$    & 65                 & 64.5 \\ \hline
$\Sigma_b -\Lambda_b$    & 194                 & 181 \\
$\Sigma_b^* -\Sigma_b$    & 19                 & 20.5 \\  \hline\hline
\end{tabular}
\caption{The hyperfine mass splittings, given in MeV, are compared with the experimental mass
differences of baryons.
The coupling strength in color-spin interaction, $v_0^{}$, is fitted from the $\Delta-N$ mass
difference and is used to determine the mass splittings of other resonances.}
\label{baryon splitting}
\end{table}

To illustrate this feature, the computed mass differences among several baryons are presented in
Table~\ref{baryon splitting} with the experimental mass splittings.
In the baryon sector, the overall strength $v_0^{}$ of the color-spin interaction is fitted from the
measured $\Delta-N$ mass splitting, which leads to $v_0^{} \sim (-199.6)^3$~MeV$^3$.
We use this value to calculate the hyperfine masses of other baryons.
For the constituent quark masses, we take the conventional values, $m_u^{} = m_d^{} = 330$~MeV,
$m_s^{} = 500$~MeV, $m_c^{} = 1500$~MeV, and $m_b^{} = 4700$~MeV.
As one can see from Table~\ref{baryon splitting}, the splittings from hyperfine masses are consistent
with the experimental mass splittings quite well.
The largest error is found in the mass difference of $\Sigma_b -\Lambda_b$.
But, even in this case, the experimental mass gap is only 13 MeV higher than the calculated
hyperfine mass gap.
Therefore, Table~\ref{baryon splitting} shows that the hyperfine mass splittings are useful to calculate
the mass splittings between baryons with different spins and different spin
configurations but with the same flavor.%
\footnote{Note that the $\Lambda$ baryon contains a spin-0 diquark while
$\Sigma$ has a spin-1 diquark.
Thus $\Lambda$ and $\Sigma$ have different spin configurations although the both have
spin-1/2~\cite{Close}.}


\begin{table}[t]
\centering
\begin{tabular}{c|c|c}  \hline\hline
              & $\Delta m$ from data & $\Delta m$ from $\langle V \rangle$ \\ \hline
$\rho-\pi$         & 635                 & 635 (fit)\\
$K^* -K$        & 396                  & 419.1 \\
$D^* -D   $     & 140                 & 139.7 \\
$D_s^* -D_s $    & 144                 & 92.2 \\
$B^* -B$    & 45.8                 & 44.6 \\
$B_s^* -B_s$    & 48.6                 & 29.4 \\
\hline\hline
\end{tabular}
\caption{The hyperfine mass splittings, given in MeV, are compared with the experimental mass
differences of mesons.
The coupling strength $v_0$ fixed from the $\rho-\pi$ mass difference is used to determine the
mass splittings of other resonances.}
\label{meson splitting}
\end{table}

Similar calculations can be performed for the meson sector and the results are given in
Table~\ref{meson splitting}.
In this case, we fit $v_0^{}$ from the observed $\rho-\pi$ mass splitting which leads to
$v_0^{} \sim (-235)^3$~MeV$^3$.
This strength is somewhat different from the one fixed in the baryon sector.
There could be various reasons for this difference.
In particular, it is often believed that the pseudoscalar mesons involved in the analysis
acquire contributions from the instanton-induced interactions.
Moreover, the pion mass calculated from Eq.~(\ref{mass}) involves the hyperfine mass
about 480 MeV, which is comparable in magnitude with the leading quark mass contribution.
This situation is rather different from the baryon case where the hyperfine masses are much
smaller than the quark mass contribution.
Nevertheless, if we use this value to calculate the hyperfine masses of the other mesons, then
the mass differences among them seem to be comparable to the experimental ones.
As one can see from Table~\ref{meson splitting}, the hyperfine masses
generate the experimental mass splittings of $K^* -K$, $D^* -D$, $B^* -B$ very well, although
the agreement is not as good for $D^*_s-D_s$ and $B_s^* -B_s$.

\section{Hyperfine masses from four-quark systems}
\label{sec:hyperfine}

In Sec.~\ref{sec:wf}, we have constructed the four-quark wave functions which are relevant for
our study on heavy meson excited states.
Depending on the symmetric aspect of the antidiquark, we come up with the following two plausible
models for the four-quark wave functions:
\begin{description}
\item[Model I]
The antidiquark is symmetric in flavor ($\bar{\bm{6}}_f$) and belongs to color state $\bm{3}_c$.
In this model, the four-quark wave functions are given by Eq.~(\ref{3 wave function}).

\item[Model II]
The antidiquark is antisymmetric in flavor ($\bm{3}_c$) and belongs to color state $\bar{\bm{6}}_c$.
In this model, the four-quark wave functions are given by Eq.~(\ref{6 wave function}).
\end{description}

The hyperfine masses of the four-quark systems are matrix elements of the hyperfine potential $V$
between these four-quark wave functions.
To explain our calculation in detail, we write the color-spin interaction for the four-quark systems as
\begin{widetext}
\begin{eqnarray}
V = v_0 \Biggl [ \lambda_1 \cdot \lambda_2 {J_1\cdot J_2 \over m_1^{} m_2^{}}
+\lambda_3 \cdot \lambda_4 {J_3 \cdot J_4 \over m_3^{} m_4^{}}
+ \lambda_1 \cdot \lambda_3{J_1\cdot J_3 \over m_1^{} m_3^{}}
+ \lambda_1 \cdot \lambda_4{J_1\cdot J_4 \over m_1^{} m_4^{} }
+ \lambda_2 \cdot \lambda_3{J_2\cdot J_3 \over m_2^{} m_3^{} }
+ \lambda_2 \cdot \lambda_4{J_2\cdot J_4 \over m_2^{} m_4^{} } \Biggr ]  ,
\label{color_spin_4_quark}
\end{eqnarray}
\end{widetext}
where the indices 1,2,3,4 refer to $Q$, $q_i$, ${\bar q}^j$, and ${\bar q}^i$ in Eqs.~(\ref{3 wave function}) and (\ref{6 wave function}).
Thus, 1,2 quarks form the diquark ($Q, q_i)$ and 3,4 quarks form the antidiquark
(${\bar q}^j, {\bar q}^i$).
The corresponding quark masses are denoted by $m_1^{}$, $m_2^{}$, $m_3^{}$, and $m_4^{}$,
respectively.
Given one specific flavor combination, one can calculate the color part and spin part separately.

\subsection{Color part}

Here we calculate the color part $\lambda_i \cdot \lambda_j$ in the potential $V$.
In the case of Model I, where the wave function is given by Eq.~(\ref{3 wave function}),
the antidiquark (namely [3,4] quarks) is in color triplet state ${\bm 3}_c$,
which restricts the diquark (namely [1,2] quarks) to be in $\bar{\bm{3}}_c$ in order to make
colorless four-quark states.
Thus, the expectation values of $\lambda_1 \cdot \lambda_2$ and $\lambda_3 \cdot \lambda_4$
can be calculated as
\begin{eqnarray}
&&\langle  \lambda_1 \cdot \lambda_2 \rangle_{\bar{\bm{3}}_c, \bm{3}_c}^{} =
\langle \lambda_3 \cdot \lambda_4  \rangle_{\bar{\bm{3}}_c, \bm{3}_c}^{} = -{8 \over 3}  .
\end{eqnarray}
In the case of Model II, where the wave function is given by Eq.~(\ref{6 wave function}),
the antidiquark is in $\bar{\bm{6}}_c$, which restricts the diquark to be in the color state ${\bf 6_c}$.
The expectation values of $\lambda_1 \cdot \lambda_2$ and $\lambda_3 \cdot \lambda_4$
can be calculated in the [1,2][3,4] basis as
\begin{eqnarray}
&&\langle \lambda_1 \cdot \lambda_2 \rangle_{{\bf 6_c}, \bar{\bm{6}}_c}^{}=
\langle \lambda_3 \cdot \lambda_4 \rangle_{{\bf 6_c}, \bar{\bm{6}}_c}^{} = {4 \over 3}  .
\end{eqnarray}

To calculate the expectation values of other operators like $\lambda_1 \cdot \lambda_3$ and
$\lambda_2 \cdot \lambda_3$, etc, it is necessary to rearrange the wave function of definite color
states in the diquark-antidiquark ([1,2][3,4]) basis into the [1,3][2,4] basis or the [1,4][2,3] basis.
This can be done by using the following decomposition
\begin{eqnarray}
q_a^{} {\bar q}^b = \underbrace{q_a^{} {\bar q}^b - {1\over 3} \delta^b_a \, q_d^{} {\bar q}^d} + \underbrace{{1\over 3} \delta^b_a \, q_d {\bar q}^d}
= {\bf 8}_a^b + \delta^b_a {\bf 1}  ,
\end{eqnarray}
which expresses a quark-antiquark pair in terms of an octet and a singlet in color space.

When the diquark and the antidiquark are in ($\bar{\bm{3}_c}, {\bm 3}_c$) as in
Eq.~(\ref{3 wave function}),  we find
\begin{eqnarray}
&& \langle  \lambda_1 \cdot \lambda_3 \rangle_{\bar{\bm{3}}_c, {\bm 3}_c}^{} =
\langle \lambda_2 \cdot \lambda_4 \rangle_{\bar{\bm{3}}_c, {\bm 3}_c}^{}
\nonumber \\
&=& \langle \lambda_2 \cdot \lambda_3 \rangle_{\bar{\bm{3}}_c, {\bm 3}_c}^{} =
\langle \lambda_1 \cdot \lambda_4 \rangle_{\bar{\bm{3}}_c, {\bm 3}_c}^{} =-{4 \over 3} .
\end{eqnarray}
\begin{widetext}
Inserting all the factors into Eq.~(\ref{color_spin_4_quark}) leads to
\begin{eqnarray}
\langle  V  \rangle_{\bar{\bm{3}}_c, {\bm 3}_c}^{} =
-{8\over 3} v_0^{} \left [ {J_1\cdot J_2 \over m_1^{} m_2^{}}
+{J_3\cdot J_4 \over m_3^{} m_4^{}} + {J_1\cdot J_3 \over 2 m_1^{} m_3^{}}
+ {J_1\cdot J_4 \over 2 m_1^{} m_4^{} } + {J_2\cdot J_3 \over 2 m_2^{} m_3^{} }
+ {J_2\cdot J_4 \over 2 m_2^{} m_4^{} } \right ] .
\label{color1}
\end{eqnarray}
When the diquark and the antidiquark are in ($\bm{6}_c, \bar{\bm{6}}_c$), the expectation values
are obtained as
\begin{eqnarray}
\langle \lambda_1 \cdot \lambda_3  \rangle_{{\bf 6_c}, \bar{\bm{6}}_c}^{} =
\langle \lambda_2 \cdot \lambda_4 \rangle_{{\bf 6_c}, \bar{\bm{6}}_c}^{}
 = \langle \lambda_2 \cdot \lambda_3 \rangle_{{\bf 6_c}, \bar{\bm{6}}_c}^{} =
\langle \lambda_1 \cdot \lambda_4 \rangle_{{\bf 6_c}, \bar{\bm{6}}_c}^{}
= - {10 \over 3}  ,
\end{eqnarray}
which leads to
\begin{eqnarray}
\langle  V  \rangle_{\bar{\bm{6}}_c, {\bf 6_c}^{} }=
{4\over 3} v_0^{} \left [ {J_1\cdot J_2 \over m_1^{} m_2^{}}
+{J_3\cdot J_4 \over m_3^{} m_4^{}}
-{5 \over 2} \left ( {J_1\cdot J_3 \over m_1^{} m_3^{}}
+ {J_1\cdot J_4 \over m_1^{} m_4^{} } +{J_2\cdot J_3 \over m_2^{} m_3^{} }
+ {J_2\cdot J_4 \over m_2^{} m_4^{} } \right ) \right ] .
\label{color2}
\end{eqnarray}
\end{widetext}

\subsection{Spin part}

The spin parts can be calculated in a  similar way.
For an illustration, we take the four-quark wave function of spin 0, which has the spin
configuration $|J, J_{12}, J_{34} \rangle = | 0 1 1 \rangle$ in the [1,2][3,4] basis.
The calculation for the other spin configurations can be done similarly.
The spin interactions, $J_1\cdot J_2$ and $J_3 \cdot J_4$, can be calculated directly on
$|0 1 1 \rangle$.
For instance, since the diquark [1,2] is in spin-1 state, $J_1\cdot J_2$ acting on $|0 1 1 \rangle$ is
\begin{equation}
J_1\cdot J_2 | 0 1 1 \rangle=\frac{1}{2}(J_{12}^2-J_1^2-J_2^2) | 0 1 1 \rangle
=\frac{1}{4} | 0 1 1 \rangle .
\end{equation}
Similarly, $J_3\cdot J_4|0 1 1 \rangle=\frac{1}{4}|0 1 1 \rangle$ since the antidiquark [3,4] is also in
the spin-1 state.

For the other spin interactions, $J_1\cdot J_3$ and $J_2\cdot J_4$ etc, it is necessary to write
the spin state $|0 1 1 \rangle$ in the [1,3][2,4] basis using Racah coefficients.
To do this, we first write $|0 1 1 \rangle$ in terms of the diquark spin and its projection
$|J_{12} M_{12} \rangle$, and the antidiquark part $|J_{34} M_{34} \rangle$, with appropriate
Clebsch-Gordan coefficients, namely,
\begin{eqnarray}
\mid 0 1 1 \rangle_{[12][34]} &=& {1\over \sqrt{3}} \Bigl [ \mid 1 1\rangle_{12} \mid 1 -1\rangle_{34}
- \mid 1 0\rangle_{12} \mid 1 0\rangle_{34}
\nonumber \\ && \mbox{} \qquad
+ \mid 1 -1\rangle_{12} \mid 1 1\rangle_{34} \Bigr ].
\label{spin wave function}
\end{eqnarray}
Here the subscripts in the kets indicate the participating quarks or antiquarks in making the
designated spin state.
Then, after writing down each spin state in terms of spinors of participating quarks,
we reorganize the $|0 1 1 \rangle$ state with respect to $|J_{13},M_{13}\rangle$ and
$|J_{24},M_{24}\rangle$.
This procedure applied to Eq.~(\ref{spin wave function}) yields the spin wave functions,
\begin{eqnarray}
\mid 0 1 1 \rangle_{[13][24]} &=& {\sqrt{3}\over 6 } \Bigl [
\mid 1 0\rangle_{13} \mid 1 0\rangle_{24} +
3 \mid 0 0\rangle_{13} \mid 0 0\rangle_{24}
\nonumber \\ && \mbox{}  \quad
- \mid 1 1\rangle_{13} \mid 1 -1\rangle_{24}
-|1 -1\rangle_{13} | 1 1\rangle_{24} \Bigr ]
\nonumber \\
\label{spin wave function 2}
\end{eqnarray}
in the [1,3][2,4] basis.
Of course, this state is not an eigenstate of $J_{13}$ as it should be.
Similarly, one can write Eq.~(\ref{spin wave function}) in terms of [1,4][2,3] spin basis,
$| J_{14},M_{14} \rangle$ and $|J_{23},M_{23}\rangle$, which gives
\begin{eqnarray}
|0 1 1 \rangle_{[14][23]} &=& {\sqrt{3}\over 6} \Bigl [ | 1 0\rangle_{14} | 1 0\rangle_{23}
+ 3| 0 0\rangle_{14} | 0 0\rangle_{23} 
\nonumber \\ && \mbox{}
-|1 1\rangle_{14} | 1 -1\rangle_{23}
-|1 -1\rangle_{14} | 1 1\rangle_{23} \Bigr ]
\label{spin wave function 3}
\end{eqnarray}
in the [1,4][2,3] basis.
Using these expressions, it is now straightforward to calculate the expectation values of the
spin operators of concern in this particular four-quark state,
$\langle 0 1 1|J_2\cdot J_4|0 1 1 \rangle$, etc.
They are obtained as
\begin{eqnarray}
&& \langle 0 1 1|J_1\cdot J_4|0 1 1 \rangle= \langle 0 1 1|J_2\cdot J_4|0 1 1 \rangle
\nonumber \\
&=& \langle 0 1 1|J_2\cdot J_3|0 1 1 \rangle=\langle 0 1 1|J_2\cdot J_4|0 1 1 \rangle=-\frac{1}{2}.
\end{eqnarray}

One interesting remark is that, under the change of basis, one can identify the decay channels of the
four-quark state of concern.
For example, in Eq.~(\ref{spin wave function 3}), the [1,4] indices correspond to $Q{\bar q} (q=u,d,s)$
and the [2,3] correspond to $q{\bar q}$.
The spin state $| 0 0\rangle_{14} | 0 0\rangle_{23}$ in Eq.~(\ref{spin wave function 3}) contains
a Fock space of pseudoscalar-pseudoscalar particles, which can decay, for instance, to $\pi D$
for $Q=c$ if the decay occurs through a `fall-apart' mechanism.
The colors of course should be combined into a singlet separately in [1,4] and [2,3] for such a decay
to happen.
The other spin states in Eq.~(\ref{spin wave function 3}) correspond to vector-vector channel like
the $\rho D^*$ channel.
From this change of spin basis, we see that the state $|0 1 1 \rangle$ consists of
pseudoscalar-pseudoscalar and vector-vector components with the probability ratio of 3:1.
Thus, this four-quark state in spin-0 channel has large component in the pseudoscalar-pseudoscalar
channel like $\pi D$.
Usually the invariant mass of this decay channel is expected to be quite lower than the possible
four-quark mass.
This means that the four-quark state with $|0 1 1 \rangle$ may have a large decay width, which
would make them difficult to be observed experimentally.
Indeed, as we mentioned in Sec.~II, $D^{*0}_0(2318)$, which is one candidate of four-quark states,
has the broad width of 267 MeV.
Also by applying the same argument to the bottom sector, we expect that $B$-meson excited
states with spin-0 are expected to be broad.
Currently, $B$-mesons with spin-0 are missing in PDG (see Table~\ref{excited}),
which might due to experimental difficulties coming from their broad widths.


\begin{table*}[t]
\centering
\begin{tabular}{c|c|l}  \hline\hline
$|J, J_{12}, J_{34} \rangle $  & ~Color of ${\bar q}_3^{} {\bar q}_4^{}$
& ~Hyperfine mass $\langle V \rangle_{q_1^{} q_2^{} {\bar q}_3^{} {\bar q}_4^{}}$  \\[1mm]
\hline
$|0 1 1 \rangle $ &  &
$\displaystyle -{2 \over 3} v_0^{} \Big[ {1 \over m_1^{} m_2^{}} + {1 \over m_3^{} m_4^{}}
- {1 \over m_1^{} m_3^{}} - {1 \over m_1^{} m_4^{}} - {1 \over m_2^{} m_3^{}} - {1 \over m_2^{} m_4^{}}
\Big] $ \\[3mm]
$|1 0 1 \rangle $ &    &
$\displaystyle -{2 \over 3 } v_0^{} \Big [ - {3 \over m_1^{} m_2^{}}+{1 \over m_3^{} m_4^{}}
\Big] $  \\[3mm]
$|1 1 1 \rangle $ & $ \bm{3}_c$ &
$\displaystyle -{2\over3 } v_0^{} \Big [ {1\over m_1^{} m_2^{}}+{1\over m_3^{} m_4^{}}
-{1\over 2 m_1^{} m_3^{}}-{1\over 2 m_1^{} m_4^{}}-{1\over 2 m_2^{} m_3^{}}
-{1\over 2 m_2^{} m_4^{}} \Big ] $\\[3mm]
$|2 1 1 \rangle $ & (Model I)  &
$\displaystyle -{2 \over3 } v_0^{} \Big [ {1 \over m_1^{} m_2^{}}+{1\over m_3^{} m_4^{}}
+{1 \over 2 m_1^{} m_3^{}} + {1 \over 2 m_1^{} m_4^{}} + {1 \over 2 m_2^{} m_3^{}} +
{1 \over 2 m_2^{} m_4^{}} \Big ] $ \\[3mm]
Mixing ($|1 0 1 \rangle, |1 1 1 \rangle $) &   &
$-\displaystyle {\sqrt{2} \over 3 } v_0^{} \Big [ -{1 \over m_1^{} m_3^{}} - {1 \over m_1^{} m_4^{}}
+ {1 \over m_2^{} m_3^{}} + {1 \over m_2^{} m_4^{}} \Big ] $
\\[3mm] \hline
$|0 1 1 \rangle $ &   &
$\displaystyle {v_0^{} \over 3} \Big [ {1\over m_1^{} m_2^{}} + {1 \over m_3^{} m_4^{}}
+ {5 \over m_1^{} m_3^{}} + {5 \over m_1^{} m_4^{}} + {5 \over m_2^{} m_3^{}}
+ {5 \over m_2^{} m_4^{}} \Big ] $ \\[3mm]
$|1 0 1 \rangle $ &   &
~$\displaystyle {v_0^{} \over 3} \Big [ -{3\over m_1^{} m_2^{}}+{1\over m_3^{} m_4^{}}
 \Big ] $ \\[3mm]
$|1 1 1 \rangle $ & $\bar{\bm{6}}_c$ &
~$\displaystyle {v_0^{} \over 6} \Big [ {2\over m_1^{} m_2^{}} + {2 \over m_3^{} m_4^{}}
+ {5 \over m_1^{} m_3^{}} + {5 \over m_1^{} m_4^{}} + {5 \over m_2^{} m_3^{}} +
{5 \over m_2^{} m_4^{}} \Big ] $ \\[3mm]
$|2 1 1 \rangle $ & (Model II) &
~$\displaystyle {v_0^{} \over 6} \Big [ {2 \over m_1^{} m_2^{}} + {2 \over m_3^{} m_4^{}}
-{5\over m_1^{} m_3^{}} - {5 \over m_1^{} m_4^{}} - {5 \over m_2^{} m_3^{}} -
{5 \over m_2^{} m_4^{}}
 \Big ] $ \\[3mm]
Mixing ($|1 0 1 \rangle, |1 1 1 \rangle $) &  &
~$\displaystyle  {5 \sqrt{2} \over 6} v_0^{}
\Big [ {1\over m_1^{} m_3^{}} +{1\over m_1^{} m_4^{}}-{1\over m_2^{} m_3^{}}
-{1 \over m_2^{} m_4^{}}
 \Big ] $ \\[3mm]
\hline\hline
\end{tabular}
\caption{The hyperfine mass $\langle V \rangle$ for a given spin configurations of the four-quark
states and the color states of the antidiquark.
Hyperfine masses presented here are for a general flavor combination,
$q_1^{} q_2^{} {\bar q}_3^{} {\bar q}_4^{}$, without including the flavor normalization.
Thus, to obtain the final hyperfine masses, one needs to combine all the flavor combination as well as
the normalization according to Eqs.~(\ref{symmetric flavor wave function}) and
(\ref{antisymmetric flavor wave function}).}
\label{hyperfine mass formula}
\end{table*}

Our prescription for evaluating the spin part can be similarly applied to the other spin states, which
include the spin-1 state with two possible configurations, $|1 0 1 \rangle$ and $|1 1 1 \rangle$,
and the spin-2 state with the configuration $|2 1 1 \rangle$.
The two configurations in $J=1$, $|1 0 1 \rangle $ and $|1 1 1 \rangle $
can mix because of nonzero mixing term $\langle 1 0 1 | V |1 1 1 \rangle $.
Therefore, one needs to diagonalize the $2 \times 2$ matrix in order to calculate physical hyperfine
masses in the spin-1 channel.

\subsection{Flavor part}

The hyperfine masses for a general flavor combination, $q_1^{} q_2^{} {\bar q}_3^{} {\bar q}_4^{}$,
are presented in Table~\ref{hyperfine mass formula} where the corresponding spin configurations
as well as color structure of the antidiquark are given.
Using these formulas, one can calculate $\langle V_D \rangle_{J}^{\bar u}$,
$\langle V_D \rangle_{J}^{\bar d}$, and $\langle V_D \rangle_{J}^{\bar s}$.
The final hyperfine masses corresponding to the states $D_{J}^{\bar u}$, $D_{J}^{\bar d}$, and
$D_{J}^{\bar s}$ can be obtained by summing over all the flavor combinations according to
Eq.~(\ref{symmetric flavor wave function}) for Model~I and
Eq.~(\ref{antisymmetric flavor wave function}) for Model~II.
To be specific, in the case of Model~I, the hyperfine masses $\langle V_D \rangle_{J}^{\bar u}$,
$\langle V_D \rangle_{J}^{\bar d}$, and $\langle V_D \rangle_{J}^{\bar s}$ are calculated
schematically as
\begin{widetext}
\begin{eqnarray}
\mbox{\bf Model I:}  && \nonumber\\
\langle V_D \rangle_{J}^{\bar u} &=& {1 \over 8}
\left [ 4 \langle V_D \rangle_{Qu{\bar u}{\bar u}}
+\langle V_D \rangle_{Qd{\bar u}{\bar d}}
+\langle V_D \rangle_{Qd{\bar d}{\bar u}}
+ \langle V_D \rangle_{Qs{\bar u}{\bar s}}
+\langle V_D \rangle_{Qs{\bar s}{\bar u}} \right ] ,
\nonumber  \\
\langle V_D \rangle_{J}^{\bar d} &=& {1 \over 8}
\left [ \langle V_D \rangle_{Qu{\bar d}{\bar u}}
+\langle V_D \rangle_{Qu{\bar u}{\bar d}}
+4\langle V_D \rangle_{Qd{\bar d}{\bar d}}
+ \langle V_D \rangle_{Qs{\bar d}{\bar s}}
+\langle V_D \rangle_{Qs{\bar s}{\bar d}} \right ] ,
\nonumber \\
\langle V_D \rangle_{J}^{\bar s} &=& {1 \over 8}
\left [ \langle V_D \rangle_{Qu{\bar s}{\bar u}}
+\langle V_D \rangle_{Qu{\bar u}{\bar s}}
+\langle V_D \rangle_{Qd{\bar s}{\bar d}}
+\langle V_D \rangle_{Qd{\bar d}{\bar s}}
+ 4\langle V_D \rangle_{Qs{\bar s}{\bar s}} \right ] .
\label{hyperfine mass sym}
\end{eqnarray}
\end{widetext}
The specified flavor combination and the associated numerical factors follow from
Eq.~(\ref{symmetric flavor wave function}).
Here each term with specified flavors, for example, the term like
$\langle V_D \rangle_{Qu{\bar u}{\bar u}}$, can obtained from the general formulas given in
Table~\ref{hyperfine mass formula} with Model~I.
The isospin symmetry requires
$\langle V_D \rangle_{J}^{\bar u} = \langle V_D \rangle_{J}^{\bar d}$.

In the case of Model II, the hyperfine masses can be calculated schematically as
\begin{widetext}
\begin{eqnarray}
\mbox{\bf Model II:}  && \nonumber\\
\langle V_D \rangle_{J}^{\bar u} &=& {1 \over 4}
\left [ \langle V_D \rangle_{Qd{\bar u}{\bar d}}
+\langle V_D \rangle_{Qd{\bar d}{\bar u}}
+ \langle V_D \rangle_{Qs{\bar u}{\bar s}}
+\langle V_D \rangle_{Qs{\bar s}{\bar u}} \right ],
\nonumber \\
\langle V_D \rangle_{J}^{\bar d} &=& {1 \over 4}
\left [ \langle V_D \rangle_{Qu{\bar d}{\bar u}}
+\langle V_D \rangle_{Qu{\bar u}{\bar d}}
+ \langle V_D \rangle_{Qs{\bar d}{\bar s}}
+\langle V_D \rangle_{Qs{\bar s}{\bar d}} \right ], \nonumber \\
\langle V_D \rangle_{J}^{\bar s} &=& {1 \over 4}
\left [ \langle V_D \rangle_{Qu{\bar s}{\bar u}}
+\langle V_D \rangle_{Qu{\bar u}{\bar s}}
+\langle V_D \rangle_{Qd{\bar s}{\bar d}}
+\langle V_D \rangle_{Qd{\bar d}{\bar s}}
 \right ] ,
\label{hyperfine mass antisym}
\end{eqnarray}
\end{widetext}
where the specified flavor combination and the numerical factors follow from
Eq.~(\ref{antisymmetric flavor wave function}).
Here each term with specified flavors is again obtained from the general formulas given
in Table~\ref{hyperfine mass formula} with Model~II.

\section{Results and discussion}
\label{sec:results}

We now present and discuss the results obtained from the two models using
Eqs.~(\ref{hyperfine mass sym}) and (\ref{hyperfine mass antisym}).
In our calculations, there are a few parameters to be fixed.
For the constituent quark masses, we use $m_u^{} = m_d^{} = 330$~MeV, $m_s^{} = 500$~MeV,
$m_c^{} = 1500$~MeV, and $m_b^{} = 4700$~MeV as discussed in Sec.~IV.

One additional parameter is the strength of the color-spin interaction $v_0^{}$.
Our analyses in Sec.~IV show that fitted parameter $v_0^{}$ takes different values
for the baryon sector and for the meson sector.
With keeping this limitation in mind, we fix $v_0^{}$ separately within our four-quark systems.
Specifically, we fix this strength from the experimental mass difference between $D^{*0}_0(2318)$
and $D^{*0}_2(2463)$ by identifying $D_{0}^{\bar u}$ with $D^{*0}_0(2318)$ and $D_{2}^{\bar u}$
with $D^{*0}_2(2463)$.
Of course, the extracted parameter $v_0^{}$ depends on the two models presented above.
Once $v_0^{}$ is fixed, one can calculate the hyperfine masses of the other resonances such as
spin-1 mesons without strangeness and spin-0,1,2 mesons with nonzero strangeness.
The obtained mass difference will be compared with the measured data to test the idea of
four-quark structure.
To check the parameter dependence of our results, we will also show the results using the
$v_0^{}$ value fixed from the $\Delta-N$ mass difference.

The calculations are performed for the charm sector and for the bottom sector.
For $B$-mesons, we will use a similar nomenclature, i.e., $B_{J}^{{\bar q}^j}$ represents a state
and $\langle V_B \rangle_{J}^{{\bar q}^j}$ (${\bar q}^j={\bar u},{\bar d},{\bar s}$) is the
corresponding hyperfine mass.

\subsection{Results from Model I}

In Model~I, the antidiquark is symmetric in flavor space and its color wave function belongs to
$\bm{3}_c$.
The hyperfine masses are calculated using Eq.~(\ref{hyperfine mass sym}).
From the mass splitting between $D^{*0}_0(2318)$ and $D^{*0}_2(2463)$ we have
$v_0^{} \sim(-193)^3$~MeV$^3$ in Model~I, which is somewhat close to the one obtained from
the $\Delta -N$ mass difference, $v_0^{} \sim (-199.6)^3$~MeV$^3$.
Using this parameter, we calculate the hyperfine masses from the four-quark states,
$|J, J_{12}, J_{34} \rangle = |0 1 1 \rangle$, $|1 0 1 \rangle$, $|1 1 1 \rangle$,
$|2 1 1 \rangle$ as well as the mixing terms between the two states in spin-1 channel.
The resulting hyperfine masses are presented in Table~\ref{hyperfine mass 1}.

We now discuss the results for the mesons without strangeness, $D_{J}^{\bar u}$, using the
corresponding hyperfine masses $\langle V_D \rangle_{J}^{\bar u}$.
In spin-1 channel, because of the two spin configurations and the mixing between them,
the hyperfine masses form a $2\times 2$ matrix.
The physical hyperfine masses can be obtained by diagonaliziation as shown below.
\begin{widetext}
\begin{eqnarray}
\begin{array}{c|lr}
& |101\rangle & |111 \rangle \\
\hline
|101 \rangle & 13.66 & 42.00\\
|111 \rangle & 42.00  & 0.82
\end{array}
\quad
&\overrightarrow{\rm diagonalization}&
\quad
\begin{array}{c|cc}
 & |D^{\bar u}_{1P}\rangle & |D^{\bar u}_{1N}\rangle \\
\hline
|D^{\bar u}_{1P}\rangle & 49.73 & 0.00\\
|D^{\bar u}_{1N}\rangle & 0.00  & -35.24
\end{array}
\end{eqnarray}
\end{widetext}
Thus, in the spin-1 channel, the physical hyperfine masses are
\begin{equation}
\langle V_D \rangle^{\bar u}_{1P}=49.73 \mbox{ MeV}, \quad
\langle V_D \rangle^{\bar u}_{1N}=-35.24 \mbox{ MeV} .
\end{equation}
Here we have denoted corresponding eigenstates as $D^{\bar u}_{1P}$ and $D^{\bar u}_{1N}$ where
the subscript $P$ ($N$) is introduced to indicate a positive (negative) hyperfine mass.

The hyperfine mass difference between $D^{\bar u}_{1P}$ and the spin-2 meson
is $\langle V_D \rangle^{\bar u}_{1P}-\langle V_D \rangle^{\bar u}_{2}=-47.48$~MeV,
which means that the $D^{\bar u}_{1P}$ mass is lower than the spin-2 meson by
about $-48$~MeV.
If we use the the experimental mass of the spin-2 meson, i.e., 2462.6 MeV, Model~I predicts the
mass of $D^{\bar u}_{1P}$ to be 2415~MeV, which is very close to the observed mass of
$D_1^0(2421)$.


\begin{table}[t]
\centering
\begin{ruledtabular}
\begin{tabular}{c|c|c|c|c}
$|J, J_{12}, J_{34} \rangle $  &   $\langle V_D \rangle_{J}^{\bar u}$ &  $\langle V_D \rangle_{J}^{\bar s}$ &  $\langle V_B \rangle_{J}^{\bar u}$  &  $\langle V_B \rangle_{J}^{\bar s}$ \\ \hline
$|0 1 1 \rangle $ &                               $-47.37$ & $-37.89$ & $-40.80$ & $-33.55$\\
$|2 1 1 \rangle $ &                               $97.21$ & $67.05$ & $84.90$ & $56.70$\\
$|1 0 1 \rangle $ &                               $13.66$ & $0.00$ & $31.71$ & $16.38$\\
$|1 1 1 \rangle $ &                               $0.82$ & $-2.91$ & $1.10$ & $-3.47$\\
mixing ($|1 0 1 \rangle, |1 1 1 \rangle $) &      $42.00$ & $29.12$ & $50.90$ & $36.05$\\
\hline
Charge                                             & $0$     & $+1$    & $-1$    & $0$    \\
\end{tabular}
\end{ruledtabular}
\caption{The hyperfine masses obtained for open-charm ($D_{J}^{\bar u}$, $D_{J}^{\bar s}$)
and open-bottom ($B_{J}^{\bar u}$, $B_{J}^{\bar s}$) excited mesons in Model~I.
Here the diquark and the antidiquark belong to the color state $\bar{\bm{3}}_c$ and $\bm{3}_c$,
respectively.
The strength of the color-spin interaction $v_0^{}$ fixed by the mass difference
$D^{*0}_2(2463)-D^{*0}_0(2318)$ is $v_0^{} \sim(-193)^3$~MeV$^3$.
We also indicate the charge of the four-quark states corresponding to the hyperfine masses.}
\label{hyperfine mass 1}
\end{table}

The other member in spin-1 channel, $D^{\bar u}_{1N}$, has a hyperfine mass of $-35.24$~MeV.
The hyperfine mass difference from the spin-2 meson is then
$\langle V_D \rangle^{\bar u}_{1N} - \langle V_D \rangle^{\bar u}_{2} = -132.45$~MeV,
which indicates that the $D^{\bar u}_{1N}$ mass should be around 2330~MeV.
The current compilation of PDG does not list the resonance corresponding to
$D^{\bar u}_{1N}$ of spin-1.
The listed $D^0_1 (2427)$ has a mass of 100 MeV larger than this estimation.
We expect that this state, if it exists, has a large decay width coming from the kinematically
favorable $\pi D^*$ mode and, therefore, it may not be easy to be identified in experiments.

This can be explained by writing the two eigenstates in spin-1 with respect to the original spin
configurations via
\begin{eqnarray}
|D^{\bar u}_{1P}\rangle &=& \alpha |101 \rangle + \beta |111 \rangle \label{mixing1} ,\\
|D^{\bar u}_{1N}\rangle &=& -\beta |101 \rangle + \alpha |111 \rangle \label{mixing2} .
\end{eqnarray}
The mixing parameters are calculated to be $\alpha=-0.76$ and $\beta=-0.65$.
Because of the sign difference in Eqs.~(\ref{mixing1}) and (\ref{mixing2}), the two spin configurations
in $J=1$ channel either add up or partially cancel in making the eigenstate $D^{\bar u}_{1P}$
or $D^{\bar u}_{1N}$.
If the two spin configurations, $|101 \rangle$ and $|111 \rangle$, are rewritten in terms of the
[1,4][2,3] basis similarly as was done in Eq.~(\ref{spin wave function 3}), one can see that they
contain the spin components, ($J_{14}=1$, $J_{23}=0$), ($J_{14}=0$, $J_{23}=1$), and
($J_{14}=1$, $J_{23}=1$).
The spin component ($J_{14}=1$, $J_{23}=0$) contains the $\pi D^*$ decay mode in addition to the
kinematically forbidden mode $K D_s^*$.
The $\pi D^*$ decay mode is kinematically favorable because the threshold energy is about 150 MeV
lower than the expected mass of $D^{\bar u}_{1N}$ which is around $2330$~MeV.
If we count only the spin part of the wave functions, the spin component containing the $\pi D^*$ mode
constitutes 25\% in the configuration $|101 \rangle$, while it is 50\% in $|111 \rangle$ before the
mixing.
After the mixing through Eqs.~(\ref{mixing1}) and (\ref{mixing2}), this component is enhanced
($\sim 74$\%) in $D^{\bar u}_{1N}$ but strongly suppressed ($\sim 0.7$\%) in $D^{\bar u}_{1P}$.
Because of strong enhancement of the component containing $\pi D^*$, $D^{\bar u}_{1N}$ is
expected to have a large decay width.
On the other hand, $D^{\bar u}_{1P}$ contains a small component containing the $\pi D^*$ mode
and is expected to be a sharp resonance.
Indeed, the $D_1^0 (2421)$, which we identify as $D^{\bar u}_{1P}$ in our model, has the decay
width about only 27 MeV.

We now discuss the results for the mesons with nonzero net strangeness, $D_{J}^{\bar s}$.
From Table~\ref{hyperfine mass 1}, we see that the hyperfine mass difference between $J=0$ and
$J=2$ channels is $\langle V_D \rangle^{\bar s}_2 - \langle V_D \rangle^{\bar s}_0 = 104$~MeV.
If we identify $D_{2}^{\bar s}$ as $D_{s2}^{*\pm}(2572)$, the spin-0 resonance $D_{0}^{\bar s}$ must
have a mass around $2470$~MeV, i.e. about 105 MeV lower than $D_{s2}^{*\pm}(2572)$.
As we have discussed in Sec.~II, the current PDG listing does not have a corresponding spin-0
resonance in this nonzero strangeness channel.
$D_{s0}^{*\pm}(2318)$ cannot be a candidate because this resonance does not belong to
$\bar{\bm{3}}_f$.
Again, the absence of this resonance may due to its large decay width, which makes
$D_{0}^{\bar s}$ difficult to be identified experimentally.
Careful inspection of Eq.~(\ref{spin wave function 3}) where the spin-0 wave function is written in the
[1,4][2,3] basis leads to that $D_{0}^{\bar s}$ contains a large component for the $K D$
decay channel, namely $|00\rangle_{14} |00\rangle_{23}$ component.
Since the $KD$ threshold energy is $2364$~MeV and is less than the expected mass of
$D_{0}^{\bar s}$, which is $2470$~MeV, the $K D$ decay channel is kinematically favorable,
which again leads to a large decay width for $D_{0}^{\bar s}$.

On the other hand, very interesting phenomena can be foreseen in spin-1 resonance
$D_{1}^{\bar s}$.
The hyperfine mass matrix for $D_{1}^{\bar s}$ in the basis of spin configurations
$|J, J_{12}, J_{34} \rangle = |1 0 1 \rangle$ and $|1 1 1 \rangle$ can be read off from
Table~\ref{hyperfine mass 1}, and its diagonalized form is as follows.
\begin{widetext}
\begin{eqnarray}
\begin{array}{c|lr}
& |101\rangle & |111 \rangle \\
\hline
|101 \rangle & 0.00 & 29.12\\
|111 \rangle & 29.12  & -2.91
\end{array}
\quad
&\overrightarrow{\rm diagonalization}&
\quad
\begin{array}{c|lr}
 & |D_{1P}^{\bar s} \rangle & |D_{1N}^{\bar s} \rangle \\
\hline
|D_{1P}^{\bar s} \rangle & 27.7 & 0.00\\
|D_{1N}^{\bar s} \rangle & 0.00  & -30.61
\end{array}
\end{eqnarray}
\end{widetext}
Thus, the physical hyperfine masses are $\langle V_D\rangle^{\bar s}_{1P}=27.7$~MeV
and $\langle V_D\rangle^{\bar s}_{1N}=-30.61$~MeV, which correspond to two spin-1 mesons
$D_{1P}^{\bar s}$ and $D_{1N}^{\bar s}$, respectively.
The two eigenstates, $D_{1P}^{\bar s}$ and $D_{1N}^{\bar s}$, are related to the original spin
configurations via
\begin{eqnarray}
|D^{\bar s}_{1P}\rangle &=& \alpha |101 \rangle + \beta |111 \rangle \label{s_mixing1} ,\\
|D^{\bar s}_{1N}\rangle &=& -\beta |101 \rangle + \alpha |111 \rangle \label{s_mixing2} ,
\end{eqnarray}
where the mixing parameters are calculated as $\alpha=-0.725$ and $\beta=-0.689$.

These two states in the spin-1 channel, $D_{1P}^{\bar s}$ and $D_{1N}^{\bar s}$, seem to fit well
with $D_{s1}^{\pm}(2535)$ and $D_{s1}^{\pm}(2460)$ of PDG.
The predicted mass of $D_{1P}^{\bar s}$, determined from the hyperfine mass difference,
$\langle V_D \rangle^{\bar s}_{1P}-\langle V_D \rangle^{\bar s}_{2}=-39$~MeV,
is $2530$~MeV.
This is very close to the observed mass, $2535$~MeV, of $D_{s1}^{\pm}$.
For $D_{1N}^{\bar s}$, the predicted mass reads about 2475~MeV, which is
only 15~MeV larger than the observed mass of $D_{s1}^{\pm} (2460)$.

One very interesting feature of this model is that $D_{1N}^{\bar s}$, which we identify as
$D_{s1}^{\pm}(2460)$, has a narrow width (see Table~\ref{excited}), while the corresponding state
in the nonstrange sector, $D^{\bar u}_{1N}$ discussed above, has a broad width.
The reason of this feature is that the possible decay channel of $D_{1N}^{\bar s}$ with the
lowest-invariant mass is kinematically forbidden.
To illustrate this, we again reorganize the spin configurations $|101\rangle$ and  $|111 \rangle$
in terms of the [1,4][2,3] basis.
Because of the nonzero strangeness, one can see that, in the case of $D_{1N}^{\bar s}$,
the decay channel with the lowest invariant mass is $K D^*$.
This is in contrast to the case of $D_{1N}^{\bar u}$ where the lowest decay channel is $\pi D^*$.
Since the $K D^*$ threshold is $\sim 2504$~MeV and is larger than the predicted mass of
$D_{1N}^{\bar s}$, that is $\sim 2474$~MeV, $D_{1N}^{\bar s}$ cannot decay into $K D^*$
even if it acquires a large $K D^*$ component from the mixing.
For $D_{1P}^{\bar s}$, its predicted mass (2533~MeV) is larger than the $K D^*$ threshold
(2504~MeV).
But in this case, the $K D^*$ component is strongly suppressed through the mixing, which again
leads to a narrow resonance.
The agreement with the experimental masses as well as the possible explanation for their decay
patterns provides strong support for the four-quark structure of excited heavy mesons.

This model can also be applied to $B$-meson systems and the results for the hyperfine masses
read
\begin{widetext}
\begin{eqnarray}
&J=0&: \quad \langle V_B\rangle^{\bar u}_{0}  = -40.8~{\rm MeV} ,
\quad \langle V_B\rangle^{\bar s}_{0} = -33.55~{\rm MeV},
\label{I_Bhy_spin0} \\
&J=1& : \quad \langle V_B\rangle^{\bar u}_{1P}  = 69.56~{\rm MeV},
\quad \langle V_B\rangle^{\bar s}_{1P} = 43.84~{\rm MeV} ,
\label{I_Bhy_spin1p} \\
&J=1 &: \quad \langle V_B\rangle^{\bar u}_{1N} = -36.75~{\rm MeV},
\quad \langle V_B\rangle^{\bar s}_{1N} = -30.94~{\rm MeV},
\label{I_Bhy_spin1n} \\
&J=2&:
\quad \langle V_B\rangle^{\bar u}_{2} = 84.9~{\rm MeV},
\quad \langle V_B\rangle^{\bar s}_{2} = 56.7~{\rm MeV} .
\label{I_Bhy_spin2}
\end{eqnarray}
\end{widetext}
We mention that the states with the superscript ${\bar u}$ have charge $-1$ and the states with the
superscript ${\bar s}$ have charge $0$.
Currently, PDG lists only 4 resonances in the excited states of $B$-mesons with relatively
well-known spin, $B_1^{0}(5724) (J=1)$, $B_2^{*0}(5743) (J=2)$, $B_{s1}^0 (5829) (J=1)$,
and $B_{s2}^{*0}(5840) (J=2)$ as can be seen in Table~\ref{excited}.
Certainly, these are not enough to test the four-quark structure.
But we can find that the four mesons listed in PDG seem to fit well with the four-quark
states $B^{\bar u}_{1P}$, $B^{\bar u}_{2}$, $B^{\bar s}_{1P}$, and $B^{\bar s}_{2}$.
The experimental mass splitting between $B_2^{*0}(5743)$ and $B_1^{0}(5724)$ is about 20~MeV
which is quite close to the corresponding value from the hyperfine mass difference
$\langle V_B\rangle^{\bar u}_{2} - \langle V_B\rangle^{\bar u}_{1P} \simeq 15$~MeV.
In $B_s$ family, the mass difference between $B_{s2}^{*0}(5840)$ and $B_{s1}^0(5829)$ is
about 10~MeV and this is again not so different from the hyperfine mass difference
$\langle V_B\rangle^{\bar s}_{2} - \langle V_B\rangle^{\bar s}_{1P} \simeq 13$~MeV.
Therefore, as far as the mass difference is concerned, the $B$-mesons in the current list of PDG
fit very well with our four-quark model.
Since the information on the $B$-meson spectroscopy is accumulating year by year in PDG at these
days, we expect that the predicted $B$-meson spectrum can be tested in near future.

The hyperfine mass differences obtained in this model are collected in
Table~\ref{heavy mesons splitting} in the column of Model~I.
Two sets of the results are shown there depending on the value of the color-spin strength $v_0^{}$.
The first set uses the $v_0^{}$ value fixed from the mass splitting between $D^{*0}_0(2318)$ and
$D^{*0}_2(2463)$ and the results are listed in the column of `$v_0$ from 4-quark'.
The other set uses the $v_0^{}$ value fitted from the $\Delta-N$ mass splitting and the results
are given in the column of `$v_0$ from $\Delta N$'.
In this calculation, we make use of the following identification of the four-quark states:
\begin{widetext}
\begin{eqnarray}
&&D^{\bar u}_{0}  = D^{*0}_0(2318),
\quad D^{\bar u}_{1P} = D^{0}_1(2421),
\quad D^{\bar u}_{2} = D_2^{*0}(2463),
\nonumber \\
&& D^{\bar s}_{1P} = D^{\pm}_{s1}(2535),
\quad D^{\bar s}_{1N} = D^{\pm}_{s1}(2460),
\quad D^{\bar s}_{2} = D_{s2}^{*\pm}(2572),
\nonumber \\
&&B^{\bar u}_{1P} = B^{0}_1(5724),
\quad B^{\bar u}_{2} = B^{*0}_2(5743),
\nonumber \\
&&B^{\bar s}_{1P} = B_{s1}^{0}(5829),
\quad B^{\bar s}_{2} = B_{s2}^{*0}(5840) .
\end{eqnarray}
\end{widetext}

Once the model parameter is fixed, we can make prediction on the masses of the unobserved
mesons of spin-0 and spin-1, i.e.,
$B^{\bar u}_{0}$, $B^{\bar s}_{0}$, $B^{\bar u}_{1N}$, and $B^{\bar s}_{1N}$.
For the $B^{\bar u}_{0}$ mass, using the fact that
$\langle V_B \rangle^{\bar u}_{1P} - \langle V_B \rangle^{\bar u}_{0} \simeq 110$~MeV
from Eqs.~(\ref{I_Bhy_spin0}) and (\ref{I_Bhy_spin1p}), the $B^{\bar u}_{0}$ mass should be
110~MeV smaller than the $B^{\bar u}_{1P}$ mass.
Since $B^{\bar u}_{1P}$ is identified as $B_1^0(5724)$, the $B^{\bar u}_{0}$ mass is expected to be
around 5613~MeV.
One can also estimate the $B^{\bar u}_{0}$ mass from the spin-2 meson $B_2^{*0}(5743)$,
which gives 5617~MeV.
Thus, the two methods give a quite consistent prediction.
We take the average value of the two values as our prediction.
In a similar way, we have
\begin{eqnarray}
J=0: && \quad B^{\bar u}_{0}~ {\rm mass} \sim  5615~{\rm MeV},
\nonumber \\ &&
\quad B^{\bar s}_{0}~ {\rm mass} \sim  5751~{\rm MeV},
\label{Bprediction_spin0} \\
J=1: && \quad B^{\bar u}_{1N}~ {\rm mass} \sim 5619~{\rm MeV},
\nonumber \\ &&
\quad B^{\bar s}_{1N}~ {\rm mass} \sim 5753~{\rm MeV}.
\label{Bprediction_spin1n}
\end{eqnarray}
The resonances with the superscript ${\bar u}$ have charge $-1$ and isospin $1/2$ (isodoublet) so
its isospin partner should appear with the same mass.
The others with the superscript ${\bar s}$ have charge 0 and they have $I=0$ (isosinglet).
We note that the $J=0$ resonances have masses quite close to their counterparts of $J=1$,
which may cause some difficulties in discovering these new resonances.
Additionally, based on a similar discussion as in $D$-meson, we expect that the three
resonances, $B^{\bar u}_{0}$, $B^{\bar s}_{0}$, and $B^{\bar u}_{1N}$ have broad widths,
which hampers the discovery of these mesons.
However, the resonance with $J=1$ of nonzero strangeness, $B^{\bar s}_{1N}$ should appear
as a sharp resonance, if exists.
\textit{Therefore, the discovery of $B^{\bar s}_{1N}$ at a mass of $\sim 5750$~MeV may be a
good probe for understanding the structure of excited heavy mesons.}


\begin{table*}[t]
\centering
\begin{tabular}{c|c|c|c|c|c}  \hline\hline
 Mass difference         & $\Delta m_{\rm expt.}$ \cite{PDG12} & \multicolumn{2}{c|}{Model I }
 & \multicolumn{2}{c}{Model II }    \\ \cline{3-6}
                                              &                        & $v_0^{}$ from 4-quark    & $v_0^{}$ from $\Delta - N$
                                              & $v_0^{}$ from 4-quark & $v_0^{}$ from $\Delta - N$   \\ \hline
$D_2^{*0} (2463) - D_0^{*0}(2318) $           & 144.6                  & 144.6 (fit)
& 160.3                      & 144.6 (fit)               & 356\\[1mm]
$D_1^0 (2421) - D_0^{*0}(2318) $              & 103.3                  & 97.1
& 107.6                      & 124.6                    & 306.7\\[1mm]
$D_2^{*0} (2463) - D_1^0(2421)   $            & 41.3                   & 47.5
& 52.6                       & 20                       &49.3\\ \hline
$D_{s2}^{*\pm}(2572) - D_{s1}^{*\pm}(2535)$   & 36.8                   & 39.4
& 43.6                       & 16.78                    &41.3\\[1mm]
$D_{s2}^{*\pm}(2572) - D_{s1}^{\pm}(2460)$    & 112.3                  & 97.7
& 108.2                      & 124.7                    &306.9\\[1mm]
$D_{s1}^{*\pm}(2535) -D_{s1}^{\pm}(2460)$     & 75.5                   & 58.3
 & 64.6                       & 107.9                    &265.6\\ \hline
$B_2^{*0}(5743) -B_1^0(5724)$                 & 19.5                   & 15.3
& 17                         & 6.98                     &16.7\\[1mm]
$B_{s2}^{*0}(5840) - B_{s1}^0(5829)$          & 10.3                   & 12.9
& 14.3                       & 5.7                      &14.0\\  \hline\hline
\end{tabular}
\caption{The mass splittings among the excited heavy mesons in MeV.
The results given under the column name `$v_0^{}$ from 4-quark' are obtained with the $v_0^{}$
value fixed from the mass difference of $D_2^{*0} (2463) - D_0^{*0}(2318)$, which gives
$v_0=(-192.9)^3$~MeV$^3$ for Model~I and $v_0=(-147.8)^3$~MeV$^3$
for Model~II.
The results given under the column name ` `$v_0$ from $\Delta N$' are obtained with the $v_0^{}$
value fixed from the $\Delta-N$ mass difference, which gives $(-199.6)^3$~MeV$^3$ in both
models.
The experimental data are from Ref.~\cite{PDG12}.}
\label{heavy mesons splitting}
\end{table*}

\subsection{Results from Model II}

Another four-quark wave function that we have constructed in Sec.~V is called Model II, where the
antidiquark is antisymmetric in flavor space and its color wave function belongs to $\bar{\bm{6}}_c$.
Within this model, the formulas for the hyperfine masses of one specific flavor combination are given
in Table~\ref{hyperfine mass formula}.
Putting them into Eq.~(\ref{hyperfine mass antisym}), we then calculate the hyperfine masses in
Model II.
Again, the strength of the color-spin interaction $v_0^{}$ is determined by fitting the mass
splitting between $D^{*0}_0(2318)$ and $D^{*0}_2(2463)$, which gives
$v_0^{} \sim(-147.8)^3$~MeV$^3$.
Using this strength, we calculate the hyperfine masses of the four-quark states,
$|J, J_{12}, J_{34} \rangle = |0 1 1 \rangle, |1 0 1 \rangle, |1 1 1 \rangle, |2 1 1 \rangle$
as well as the mixing term between the two spin-1 states.
Again for spin-1 case, it is necessary to diagonalize the hyperfine masses in order to obtain the
physical states.

Then we can make predictions on the excited heavy meson spectrum as we did for Model~I.
The hyperfine masses for $D$ and $D_s$ family are obtained as
\begin{widetext}
\begin{eqnarray}
&J=0&: \quad \langle V_D\rangle^{\bar u}_{0}  = -106.43~{\rm MeV},
\quad \langle V_D\rangle^{\bar s}_{0} = -108.82~{\rm MeV},
\label{II_Dhy_spin0} \\
&J=1&: \quad \langle V_D\rangle^{\bar u}_{1P}  = 18.18~{\rm MeV},
\quad \langle V_D\rangle^{\bar s}_{1P} = 24.58~{\rm MeV},
\label{II_Dhy_spin1p}\\
&J=1&:\quad \langle V_D\rangle^{\bar u}_{1N} = -79.2~{\rm MeV},
\quad \langle V_D\rangle^{\bar s}_{1N} = -83.34~{\rm MeV},
\label{II_Dhy_spin1n}\\
&J=2&: \quad \langle V_D\rangle^{\bar u}_{2} = 38.2~{\rm MeV},
\quad \langle V_D\rangle^{\bar s}_{2} = 41.36~{\rm MeV} .
\label{II_Dhy_spin2}
\end{eqnarray}
The hyperfine masses for $B$ and $B_s$ family in Model~II read
\begin{eqnarray}
&J=0&: \quad \langle V_B\rangle^{\bar u}_{0}  = -91.65~{\rm MeV},
\quad \langle V_B\rangle^{\bar s}_{0} = -95.05~{\rm MeV},
\label{II_Bhy_spin0}\\
&J=1&: \quad \langle V_B\rangle^{\bar u}_{1P}  = 25.87~{\rm MeV},
\quad \langle V_B\rangle^{\bar s}_{1P} = 31.0~{\rm MeV},
\label{II_Bhy_spin1p}\\
&J=1&: \quad \langle V_B\rangle^{\bar u}_{1N} = -82.57~{\rm MeV},
\quad \langle V_B\rangle^{\bar s}_{1N} = -86.58~{\rm MeV},
\label{II_Bhy_spin1n}\\
&J=2&: \quad \langle V_B\rangle^{\bar u}_{2} = 32.65~{\rm MeV},
\quad \langle V_B\rangle^{\bar s}_{2} = 36.69~{\rm MeV} .
\label{II_Bhy_spin2}
\end{eqnarray}
\end{widetext}

Alternatively, within Model II, we can again calculate the mass differences by using the $v_0^{}$
value determined by the $\Delta-N$ mass difference.
Presented in Table~\ref{heavy mesons splitting} are the mass differences in Model~II for these
two values of $v_0^{}$.
These results are compared with the experimental mass splittings as well as the predictions of
Model~I.
As one can see in Table~\ref{heavy mesons splitting}, the results from Model~I have a better
agreement with the experimental data than those of Model II.
Therefore, we conclude that the four-quark wave functions constructed in Model I are more
reliable for the excited heavy meson states as far as the mass differences are concerned.

\section{Summary}
\label{sec:summary}

In this work, we have constructed four-quark wave functions, which might be relevant for
excited states of open charm and open bottom mesons.
The four-quark wave functions were constructed from a diquark-antidiquark picture
under the assumption that they form the $\bar{\bm{3}}_f$ multiplet in the SU(3) flavor space.
Formation of $\bar{\bm{3}}_f$ seems to be realized in some of the observed excited states.
Within this approach, we propose two models for the four-quark wave functions, which we call
Model~I and Model~II.
In Model~I, the antidiquark is symmetric in flavor (${\bar{\bm{6}}_f}$) and antisymmetric in color
(${\bm 3}_c$).
On the contrary, in Model II, the antidiquark is antisymmetric in flavor (${\bm 3}_f$) and symmetric
in color (${\bar{\bm{6}}_c}$).
In both models, the possible spin structures are found to be
$|J,J_{12},J_{34} \rangle = |011 \rangle$, $|101 \rangle$, $|111 \rangle$, and $|211 \rangle$
where $J$ is the spin of the four-quark system, $J_{12}$ the diquark spin,
$J_{34}$ the antidiquark spin.
There exists a mixing between the two spin-1 states, which is to be diagonalized for finding the
physical states.
To test these four-quark structure, we calculated the hyperfine masses using the color-spin
interactions and investigated whether they can reproduce the observed mass splittings among the
excited states of $D$, $D_s$, $B$ and $B_s$ families listed in PDG.

By comparing with the experimental masses, we found that Model~I gives a good description of the
observed mass splittings as shown in Table~\ref{heavy mesons splitting} while Model II fails.
It should be noted that all these results are obtained with only one model parameter $v_0^{}$ which
is fixed either by the mass splitting between $D^{*0}_0(2318)$ and $D^{*0}_2(2463)$
or by the $\Delta-N$ mass splitting.
We found that Model~I gives a nice description of the mass splittings with these two values
of $v_0^{}$.

Another supporting result of four-quark structure is the appearance of two spin-1 states.
This is indeed consistent with the two experimentally observed resonances,
$D_{s1}^{\pm}(2460)$ and $D_{s1}^{*\pm}(2535)$, of which masses are well explained
by our Model~I.
On the other hand, in the charm sector, one of the two spin-1 states fits nicely with the
$D_1^0(2421)$ meson but there is a missing resonance.
We have demonstrated that the missing spin-1 state may have a large component of
$\pi D^*$ decay mode which is substantially magnified through the mixing.
Because of this decay channel, this resonance is expected to be a broad resonance and
it may not be easily identified in experiments.
However, the two states in $D_s$ mesons have smaller decay widths.
In this case, the decay mode with the lowest invariant mass is $K D^*$ which is kinematically
forbidden in one state and, in the other state, this decay mode is strongly suppressed through
the mixing.

Our Model I can predict some other resonances which are currently missing in PDG compilation.
Motivated by tits success to explain the observed psectroscopy, we make predictions on some
missing resonances as follows.
\begin{eqnarray}
&J=0&: \quad D^{\bar s}_0 \sim 2468~{\rm MeV} ; \quad \mbox{broad resonance}
\nonumber ,\\
&J=1&: \quad D^{\bar u}_{1N}, D^{\bar d}_{1N} \sim 2330~{\rm MeV} ;\quad
\mbox{broad resonances} ,
\nonumber \\
&J=0&: \quad B^{\bar u}_{0}, B^{\bar d}_{0} \sim  5615~{\rm MeV} ;\quad
\mbox{broad resonances}, \nonumber \\
&J=0&: \quad B^{\bar s}_{0} \sim  5751~{\rm MeV} ; \quad \mbox{broad resonance},
\nonumber \\
&J=1&: \quad B^{\bar u}_{1N}, B^{\bar d}_{1N} \sim 5619~{\rm MeV} ;\quad
\mbox{broad resonance},
\nonumber \\
&J=1&: \quad  B^{\bar s}_{1N} \sim 5753~{\rm MeV} ;\quad \mbox{narrow resonance}.
\end{eqnarray}
This shows that most of these resonances are expected to have broad widths
due to decay modes kinematically allowed.
Therefore, those resonances may not be easily identified in experiments.
However, there is one exception: $B^{\bar s}_{1N}$ of spin-1 is expected to be a narrow resonance
because its possible decay mode $KB^*$ is not kinematically allowed.
So the discovery of $B^{\bar s}_{1N}$(5753) in future experiments will shed light on our
understanding of four-quark structure of excited heavy mesons.

Throughout the present work, our discussions are limited to the masses of resonances based on
the group structure of four-quark systems.
Then the next question would be the dynamical origin of such a structure, which may also
provide a key to understand the reason why Model~I is better than Model~II for explaining
heavy meson excited states in four-quark picture.
It is, therefore, highly desirable to test the four-quark picture based on dynamical model approaches
to calculate full mass spectra and the couplings of meson resonances.
Such studies should also address the question whether the real physical states would be mixtures
of orbitally excited two-quark states and four-quark states.
Testing the four-quark interpolating fields in QCD sum rules may also be interesting to compute
the physical properties of excited heavy mesons and it will help us verify which structure has a strong
overlap with the physical hadron states.

\acknowledgments

\newblock
We are grateful to Suhoung Lee for fruitful discussions.
The work of M.-K.C. was supported in part by the National Research Foundation of Korea under
Grant Nos. NRF-2014R1A2A2A05003548 and NRF-2012M7A1A2055605.
Y.O. was supported in part by the National Research Foundation of Korea
under Grant Nos.\ NRF-2011-220-C00011 and NRF-2013R1A1A2A10007294.


\begin{thebibliography}{10}

\bibitem{Belle03}
Belle Collaboration, S.~K. Choi \textit{et~al.\/},
\newblock Phys. Rev. Lett. \textbf{91}, 262001 (2003).

\bibitem{BABAR05b}
BaBar Collaboration, B.~Aubert \textit{et~al.\/},
\newblock Phys. Rev. Lett. \textbf{95}, 142001 (2005).

\bibitem{Belle08b}
Belle Collaboration, S.~K. Choi \textit{et~al.\/},
\newblock Phys. Rev. Lett. \textbf{100}, 142001 (2008).

\bibitem{Olsen09}
S.~L. Olsen,
\newblock Nucl. Phys. A \textbf{827}, 53c (2009).

\bibitem{MPPR14}
L.~Maiani, F.~Piccinini, A.~Polosa, and V.~Riquer,
\newblock Phys. Rev. D \textbf{89}, 114010 (2014).

\bibitem{LHCb14}
LHCb Collaboration, R.~Aaij \textit{et~al.\/},
\newblock Phys. Rev. Lett. \textbf{112}, 222002 (2014).

\bibitem{LEPS03}
LEPS Collaboration, T.~Nakano \textit{et~al.\/},
\newblock Phys. Rev. Lett. \textbf{91}, 012002 (2003).

\bibitem{Smith02}
E.~S. Smith,
\newblock Heavy Ion Phys. \textbf{16}, 187 (2002).

\bibitem{PDG12}
Particle Data Group, J.~Beringer \textit{et~al.\/},
\newblock Phys. Rev. D \textbf{86}, 010001 (2012),
\newblock http://pdg.lbl.gov.

\bibitem{Jaffe77a}
R.~L. Jaffe,
\newblock Phys. Rev. D \textbf{15}, 267 (1977).

\bibitem{Jaffe77b}
R.~L. Jaffe,
\newblock Phys. Rev. D \textbf{15}, 281 (1977).

\bibitem{Jaffe04}
R.~L. Jaffe,
\newblock Phys. Rep. \textbf{409}, 1 (2005).

\bibitem{Torn95}
N.~A. T{\"o}rnqvist,
\newblock Z. Phys. C \textbf{68}, 647 (1995).

\bibitem{LKO13}
H.-J. Lee, N.~I. Kochelev, and Y.~Oh,
\newblock Phys. Rev. D \textbf{87}, 117901 (2013).

\bibitem{MPPR04a}
L.~Maiani, F.~Piccinini, A.~D. Polosa, and V.~Riquer,
\newblock Phys. Rev. Lett. \textbf{93}, 212002 (2004).

\bibitem{EFG09}
D.~Ebert, R.~Faustov, and V.~Galkin,
\newblock Eur. Phys. J. C \textbf{60}, 273 (2009).

\bibitem{EFG11}
D.~Ebert, R.~Faustov, and V.~Galkin,
\newblock Phys. Lett. B \textbf{696}, 241 (2011).

\bibitem{Jova07}
V.~B. Jovanovic,
\newblock Phys. Rev. D \textbf{76}, 105011 (2007).

\bibitem{Dmit05}
V.~Dmitra\v{s}inovi\'c,
\newblock Phys. Rev. Lett. \textbf{94}, 162002 (2005).

\bibitem{BLMNN05}
M.~E. Bracco, A.~Lozea, R.~D. Matheus, F.~S. Navarra, and M.~Nielsen,
\newblock Phys. Lett. B \textbf{624}, 217 (2005).

\bibitem{KO05}
H.~Kim and Y.~Oh,
\newblock Phys. Rev. D \textbf{72}, 074012 (2005).

\bibitem{VR03}
E.~van Beveren and G.~Rupp,
\newblock Phys. Rev. Lett. \textbf{91}, 012003 (2003).

\bibitem{VFV06}
J.~Vijande, F.~Fernandez, and A.~Valcarce,
\newblock Phys. Rev. D \textbf{73}, 034002 (2006),
\newblock \textbf{74}, 059903(E) (2006).

\bibitem{Close}
F.~E. Close,
\newblock \textit{An Introduction to Quarks and Partons} 
(Academic Press, London, 1979).

\bibitem{APEFL93}
M.~Anselmino, E.~Predazzi, S.~Ekelin, S.~Fredriksson, and D.~B. Lichtenberg,
\newblock Rev. Mod. Phys. \textbf{65}, 1199 (1993).

\bibitem{LY09}
S.~H. Lee and S.~Yasui,
\newblock Eur. Phys. J. C \textbf{64}, 283 (2009).

\bibitem{Keren07}
B.~Keren-Zur,
\newblock Ann. Phys. (N.Y.) \textbf{323}, 631 (2008).

\bibitem{OT89}
M.~Oka and S.~Takeuchi,
\newblock Phys. Rev. Lett. \textbf{63}, 1780 (1989).

\bibitem{Silve92}
B.~Silvestre-Brac,
\newblock Phys. Rev. D \textbf{46}, 2179 (1992).

\bibitem{GR81}
S.~Gasiorowicz and J.~L. Rosner,
\newblock Amer. J. Phys. \textbf{49}, 954 (1981).

\end{thebibliography}
\end{document}